\documentclass[aps,prd,twocolumn,10pt,groupedaddress]{revtex4-1}
\usepackage{amssymb}
\usepackage{graphicx}
\usepackage{amsmath}
\usepackage{hyperref}
\usepackage{color}

\begin{document}

\title{Probing cosmic anisotropy with gravitational waves as standard sirens}
\author{Rong-Gen Cai}
\email{cairg@itp.ac.cn}
\affiliation{CAS Key Laboratory of Theoretical Physics, Institute of Theoretical Physics, Chinese Academy of Sciences, Beijing 100190, China}
\affiliation{School of Physical Sciences, University of Chinese Academy of Sciences, Number 19A Yuquan Road, Beijing 100049, China}

\author{Tong-Bo Liu}
\email{liutongbo@itp.ac.cn}
\affiliation{CAS Key Laboratory of Theoretical Physics, Institute of Theoretical Physics, Chinese Academy of Sciences, Beijing 100190, China}
\affiliation{School of Physical Sciences, University of Chinese Academy of Sciences, Number 19A Yuquan Road, Beijing 100049, China}

\author{Xue-Wen Liu}
\email{liuxuewen14@itp.ac.cn}
\affiliation{CAS Key Laboratory of Theoretical Physics, Institute of Theoretical Physics, Chinese Academy of Sciences, Beijing 100190, China}
\affiliation{School of Physical Sciences, University of Chinese Academy of Sciences, Number 19A Yuquan Road, Beijing 100049, China}

\author{Shao-Jiang Wang}
\email{schwang@itp.ac.cn}
\affiliation{CAS Key Laboratory of Theoretical Physics, Institute of Theoretical Physics, Chinese Academy of Sciences, Beijing 100190, China}
\affiliation{Department of Physics and Helsinki Institute of Physics, PL 64, FI-00014 University of Helsinki, Helsinki, Finland}
\affiliation{School of Physical Sciences, University of Chinese Academy of Sciences, Number 19A Yuquan Road, Beijing 100049, China}

\author{Tao Yang}
\email{yangtao2017@bnu.edu.cn}
\affiliation{Department of Astronomy, Beijing Normal University, Beijing, 100875, China}

\date{\today}

\begin{abstract}
  The gravitational wave (GW) as a standard siren directly determines the luminosity distance from the gravitational waveform without reference to the specific cosmological model, of which the redshift can be obtained separately by means of the electromagnetic counterpart like GW events from binary neutron stars and massive black hole binaries (MBHBs). To see to what extent the standard siren can reproduce the presumed dipole anisotropy written in the simulated data of standard siren events from typical configurations of GW detectors, we find that (1) for the Laser Interferometer Space Antenna with different MBHB models during five-year observations, the cosmic isotropy can be ruled out at $3\sigma$ confidence level (C.L.) and the dipole direction can be constrained roughly around $20\%$ at $2\sigma$ C.L., as long as the dipole amplitude is larger than $0.03$, $0.06$ and $0.025$ for MBHB models Q3d, pop III and Q3nod with increasing constraining ability, respectively; (2) for Einstein Telescope with no less than $200$ standard siren events, the cosmic isotropy can be ruled out at $3\sigma$ C.L. if the dipole amplitude is larger than $0.06$, and the dipole direction can be constrained within $20\%$ at $3\sigma$ C.L. if the dipole amplitude is near $0.1$; (3) for the Deci-Hertz Interferometer Gravitational wave Observatory with no less than $100$ standard siren events, the cosmic isotropy can be ruled out at $3\sigma$ C.L. for dipole amplitude larger than $0.03$ , and the dipole direction can even be constrained within $10\%$ at $3\sigma$ C.L. if the dipole amplitude is larger than $0.07$. Our work manifests the promising perspective of the constraint ability on the cosmic anisotropy from the standard siren approach.
\end{abstract}
\maketitle

\section{Introduction}

The cosmological principle states that our Universe is spatially homogeneous and isotropic on sufficiently large scales, which mathematically leads to the Friedmann-Lema\^{\i}tre-Robertson-Walker (FLRW) metric. Based on the FLRW metric, the modern cosmology is constructed as the $\Lambda$-cold-dark-matter ($\Lambda$CDM) model, which is consistent with various observational constraints, such as observations from cosmic microwave background (CMB) radiation \cite{Komatsu:2010fb,Hinshaw:2012aka,Ade:2013zuv,Ade:2015xua} and observations from large scale structures \cite{Peacock:2001gs,Blake:2011rj,Reid:2012sw}. However, there are still some puzzling conflicts between the $\Lambda$CDM model and some observations \cite{Perivolaropoulos:2008ud}, for example, the hemispherical power asymmetry of various CMB anomalies \cite{Schwarz:2015cma}, which was first noticed in Wilkinson microwave anisotropy probe (WMAP) data analysis \cite{Eriksen:2003db,Hansen:2004mj,Eriksen:2007pc,Hoftuft:2009rq} and later confirmed in Planck 2013/2015 data analysis \cite{Ade:2013nlj,Akrami:2014eta,Ade:2015hxq}. This hemispherical asymmetry reflects an asymmetric power of one ecliptic hemisphere with respect to the other one, which can be modeled as dipole modulation anisotropy \cite{Gordon:2005ai,Prunet:2004zy}. In Planck 2013 \cite{Ade:2013nlj}, the dipole amplitude is constrained around $0.078\pm0.021$ and the dipole direction is constrained around $(l,b)=(227^\circ,-15^\circ)\pm19^\circ$. In Planck 2015 \cite{Ade:2015hxq}, the dipole amplitude is constrained around $0.066\pm0.021$ and the dipole direction is constrained around $(l,b)=(230^\circ,-16^\circ)\pm24^\circ$. Besides the temperature data of CMB observations, this dipole anisotropy also manifests itself in polarized data \cite{Namjoo:2014pqa,Aluri:2017cna}. Apart from CMB observations, this dipole anisotropy can be seen in other cosmological observations like large scale structures \cite{Fernandez-Cobos:2013fda,Bengaly:2016amk} as well. For example, \cite{Migkas:2017vir} introduced a new method to investigate the possibility of a cosmic anisotropy, through the luminosity distance that enters via the x-ray flux-luminosity conversion, while the temperature measurement is cosmology independent. They indeed identified a cosmic dipole. For theoretical interpretations on cosmic anisotropy that have been extensively studied in previous literature, we want to mention that the cosmic anisotropies are naturally generated in the vector-tensor theories of gravity \cite{Heisenberg:2014rta,Jimenez:2016isa,Heisenberg:2016wtr}. Even if the background is chosen to be spatially homogeneous and isotropic, the vector perturbations will introduce interesting features.

Another way to probe the dipole anisotropy is to use the type Ia supernovae (SNe Ia) data. Since SNe Ia have approximately the same absolute magnitude, they can be used to measure the luminosity distance, which leads to the discovery that the expansion of the Universe is accelerating. What is more, Tsagas \cite{Tsagas:2011wq} argued that peculiar velocities introduce a preferred spatial direction, so that one may find that the acceleration is maximized in one direction and minimized in the opposite, which might be associated with dipolelike anisotropy.  Assuming the deviation of luminosity distance in isotropic background is the dipole form, Cai \textit{et al}. argued that the dipole modulation is needed at 2$\sigma$ confidence level (C.L.) and some preferred directions have been identified by using Union2 SNe Ia data and gamma-ray burst (GRB) data \cite{Cai:2011xs,Cai:2013lja}. Other studies on cosmological isotropy with the help of SNe Ia data sets can be found in \cite{Bengaly:2015dza,Lin:2015rza,Wang:2014vqa,Yang:2013gea}.

Recently, the first detection of a standard siren event from the GW170817 merger event of binary neutron stars (BNSs) by the LIGO-Virgo detector network \cite{TheLIGOScientific:2017qsa} shows us the use of GW as a standard siren \cite{Schutz:1986gp,Holz:2005df} to directly determine the luminosity distance from the gravitational waveform of coalescing binaries. With the identification of the associated electromagnetic (EM) counterpart, the redshift can also be determined. Apart from the coalescing binaries involving neutron stars, the massive black hole binaries (MBHBs) from $10^{4}$ to $10^{7}M_{\odot}$ are also expected to produce a detectable EM counterpart, because they are supposed to merge in a gas-rich environment and within the laser interferometer space antenna (LISA) frequency band \cite{Barausse:2012fy,Tamanini:2016zlh}. In addition, the MBHB standard siren will probe the cosmic expansion at distances up to $z\sim 15$ that SNIa cannot reach. So long as the luminosity distance-redshift relation is obtained, we would be able to take GW as an alternative way to probe the anisotropy of cosmic expansion.

Although, the standard siren method was used in \cite{Nishizawa:2010xx} to probe the evolution of the Hubble parameter by a dipole induced from local  peculiar velocity, there has been no study to our knowledge on the dipole anisotropy from the standard siren approach. In this paper, we fill in this gap as an application of the standard siren on cosmology \cite{Cai:2017cbj}. We first generate an anisotropic sample of standard siren events with presumed dipole field, then we populate the anisotropic sample within the detection configurations from LISA, Einstein Telescope (ET) and Deci-Hertz Interferometer Gravitational wave Observatory (DECIGO), next we apply the Markov chain Monte Carlo (MCMC) analysis to constrain this populated anisotropic sample; finally we can see to what extent these standard siren events can reproduce the presumed dipole anisotropy written in the simulated data.

The paper is organized as follows. In Sec. \ref{sec:GWsiren}, we introduce the idea about using GW as a standard siren and obtain the information of luminosity distance. In Sec. \ref{sec:simulations}, we simulate the standard siren events, introduce the dipole modulation and adopt the MCMC approach to constrain the anisotropic amplitude and direction. In Sec. \ref{sec:results}, we present our main results and discuss the constraint ability. Conclusions will be given in Sec. \ref{sec:conclusion}. Throughout this paper, a flat universe is assumed for simplicity, and the geometric unit $c=G=1$ is adopted so that $1\,\mathrm{Mpc}=1.02938\times10^{14}\,\mathrm{Hz}^{-1}$ and $1\,M_\odot=4.92535\times10^{-6}\,\mathrm{Hz}^{-1}$.

\section{Gravitational wave as  a standard siren}\label{sec:GWsiren}

In the spatially flat FLRW universe, the luminosity distance is given by
\begin{align}\label{eq:dL0}
d_{\rm L}^{0}(z)=\frac{1+z}{H_{0}}\int_{0}^{z}\frac{dz^{\prime}}{E(z^{\prime})},
\end{align}
where $E(z)\equiv H(z)/H_{0}$ and $H_{0}=100h\,\rm km\,s^{-1}\,Mpc^{-1}$ is the Hubble constant. We take the standard $\Lambda$CDM model as the fiducial cosmological model and the Hubble parameter is given by
\begin{align}\label{eq:Hubblepara}
H^{2}(z)=H_{0}^{2}[(1-\Omega_{m})+\Omega_{m}(1+z)^{3}],
\end{align}
where $\Omega_{m}$ is the matter density parameter today. We take the fiducial values
\begin{align}\label{eq:fiducialpara}
\Omega_{m}=0.308,\quad h=0.678,\quad \Omega_{K}=0
\end{align}
from the current Planck 2015 data \cite{Ade:2015xua}.

The detector response to a GW signal in the transverse-traceless (TT) gauge is given by
\begin{align}
h(t)=F_{+}(\theta,\phi,\psi)h_{+}(t)+F_{\times}(\theta,\phi,\psi)h_{\times}(t),
\end{align}
where $F_{+,\times}$ are the antenna pattern functions for the two polarizations, $h_{+}=h_{xx}=-h_{yy}, h_{\times}=h_{xy}=h_{yx}$, and ($\theta,\phi$) are the angles which denote the direction of the source in the detector frame, $\psi$ is the polarization angle. The pattern functions could be different depending on the angle spanned by the two interferometer arms.

Let us firstly consider laser interferometer space antenna (LISA), which is a space-based GW detector designed for signals of merging massive black holes, stellar black hole binaries and extreme mass ratio inspirals. We adopt the LISA configuration to be N2A2M5L6, which means 2 Gm arm length, two active laser links (arms) fixed, five-year's mission duration and six links in total. This configuration is in agreement with the first results from the LISA pathfinder mission, which also considers the actual LISA configuration that will be proposed at the L3 European Space Agency (ESA) call for mission. The pattern functions for LISA are given by \cite{Klein:2015hvg}
\begin{align}\label{eq:LISAPattern}
F_{+}^{(1)}(\theta,\phi,\psi)=&\frac{\sqrt{3}}{2}\bigg[\frac12(1+\cos^{2}(\theta))\cos(2\phi)\cos(2\psi) \notag\\
&-\cos(\theta)\sin(2\phi)\sin(2\psi)\bigg];\notag\\
F_{\times}^{(1)}(\theta,\phi,\psi)=&\frac{\sqrt{3}}{2}\bigg[\frac12(1+\cos^{2}(\theta))\cos(2\phi)\sin(2\psi)\notag \\
&+\cos(\theta)\sin(2\phi)\cos(2\psi)\bigg],
\end{align}
and the two other pattern functions are $F_{+,\times}^{(2)}(\theta,\phi,\psi)=F_{+,\times}^{(1)}(\theta,\phi-\pi/4,\psi)$.

Next we focus on the Einstein Telescope (ET), which is a proposed third-generation ground-based GW detector. The three 10-km-long arms of ET will be in an equilateral triangle, and the frequency it will cover ranges from 1 to $10^{4}$ Hz. The corresponding pattern functions for ET \cite{Zhao:2010sz} are
\begin{align}\label{eq:ETPattern}
F_{+}^{(1)}(\theta,\phi,\psi)=&\frac{\sqrt{3}}{2}\bigg[\frac12(1+\cos^{2}(\theta))\cos(2\phi)\cos(2\psi) \notag\\
&-\cos(\theta)\sin(2\phi)\sin(2\psi)\bigg];\notag\\
F_{\times}^{(1)}(\theta,\phi,\psi)=&\frac{\sqrt{3}}{2}\bigg[\frac12(1+\cos^{2}(\theta))\cos(2\phi)\sin(2\psi)\notag \\
&+\cos(\theta)\sin(2\phi)\cos(2\psi)\bigg],
\end{align}
and the rest of the pattern functions are $F_{+,\times}^{(2)}(\theta,\phi,\psi)=F_{+,\times}^{(1)}(\theta,\phi+2\pi/3,\psi)$ and $F_{+,\times}^{(3)}(\theta,\phi,\psi)=F_{+,\times}^{(1)}(\theta,\phi+4\pi/3,\psi)$, respectively, since the three interferometers align with an angle $60^\circ$ with each other.

The third detector we consider here is the Deci-Hertz Interferometer Gravitational wave Observatory (DECIGO) \cite{Kawamura:2006up}, which is a future plan of Japanese space mission for observing GWs around 0.1-10 Hz, similar to the big bang observer (BBO) proposed by America. DECIGO is made up of four trianglelike units, and for each unit, in order to obtain the orthogonal data streams, we take linear combinations so that they form an orthogonal basis for L-shaped interferometers on the detector plane; see \cite{Yagi:2011wg} for more details. For DECIGO, the antenna pattern functions are given by
\begin{align}\label{eq:DECIGOPattern}
F_{+}^{(1)}(\theta,\phi,\psi)=&\frac12(1+\cos^{2}(\theta))\cos(2\phi)\cos(2\psi)\notag\\
&-\cos(\theta)\sin(2\phi)\sin(2\psi);\notag\\
F_{\times}^{(1)}(\theta,\phi,\psi)=&\frac12(1+\cos^{2}(\theta))\cos(2\phi)\sin(2\psi)\notag\\
&+\cos(\theta)\sin(2\phi)\cos(2\psi),
\end{align}
and another pair is $F_{+,\times}^{(2)}(\theta,\phi,\psi)=F_{+,\times}^{(1)}(\theta,\phi-\pi/4,\psi)$, due to the fact that the triangle unit can be effectively regarded as the two L-shaped interferometers, and they align with the angle $45^\circ$ with each other.

Next we compute the Fourier transform of the GW signal by applying the stationary phase approximation,
\begin{align}\label{eq:FTWaveform}
\mathcal{H}(f)=\mathcal{A}f^{-7/6}e^{i\Psi(f)},
\end{align}
where the amplitude is given by
\begin{align}\label{eq:Amplitude}
\mathcal{A}=&\frac{1}{d_{\rm L}}\sqrt{F_{+}^{2}(1+\cos^{2}(\iota))^{2}+4F_{\times}^{2}\cos^{2}(\iota)}\notag\\
&\times \sqrt{5\pi/96}\pi^{-7/6}\mathcal{M}_{c}^{5/6},
\end{align}
where $\iota$ is the inclination angle between the orbit and the line of sight. The phase in Eq. \eqref{eq:FTWaveform} is computed in the post-Newtonian formalism up to 3.5 PN and the specific expression of which can be found in \cite{Nishizawa:2010xx}, therefore we can neglect the spin effects when considering the binary system. Here $\mathcal{M}_{c}$ denotes the observed total mass $\mathcal{M}_{c}=(1+z)M\eta^{3/5}$, where $M=m_{1}+m_{2}$ represents the total mass of the binary components, and $\eta=m_{1}m_{2}/M^{2}$ is the symmetric mass ratio.

One way to obtain the corresponding redshifts of GW events is to identify their electromagnetic counterparts, and it has achieved great success in GW170817 \cite{TheLIGOScientific:2017qsa}. One of the EM counterparts is short gamma-ray bursts (SGRBs), which last less than about two seconds, and they are believed to originate from the merger of two compact stars, such as BNSs. SGRBs are highly relativistic and strongly beamed phenomena, so only those propagating directly along the line of sight are likely to be detected, thus they naturally break the distance-inclination degeneracy appearing in the gravitational waveform. It is worth noting that the maximal inclination is $\iota\simeq20^\circ$ \cite{Li:2013lza}. But when we average the Fisher matrix over the inclination $\iota$ and the polarization angle $\psi$ with $\iota<20^\circ$, the result is almost the same as we take $\iota$ to be 0. In the following simulations, we therefore take $\iota=0^\circ$ for simplicity, which makes the amplitude $\mathcal{A}$ of the gravitational waveform independent of $\psi$ as well.

Note that since the observation of  the inspirals of MBHBs by LISA will last several months, the sky position of which relative to the detector will change during this time, causing a varying detector response. Therefore, we need to change the analysis of LISA slightly by taking the rotation effect into account according to \cite{Cutler:1997ta}, which is different from that of the ground-based detectors. We take the sun as the center, and modulate our simulated angles as follows:
\begin{align}\label{eq:thetaS}
\cos\theta_{\rm S}(t)=\frac{1}{2}\cos\bar{\theta}_{\rm S}-\frac{\sqrt{3}}{2}\sin\bar{\theta}_{\rm S}\cos(\bar{\phi}(t)-\bar{\phi}_{\rm S}),
\end{align}
\begin{align}\label{eq:phiS}
\phi_{\rm S}(t)&=\alpha_{1}(t)+\pi/12\notag\\
&-\tan^{-1}\bigg[\frac{\sqrt{3}\cos\bar{\theta}_{\rm S}+\sin\bar{\theta}_{\rm S}\cos(\bar{\phi}(t)-\bar{\phi}_{\rm S})}{2\sin\bar{\theta}_{\rm S}\cos(\bar{\phi}(t)-\bar{\phi}_{\rm S})}\bigg],
\end{align}
where $\bar{\theta}_{\rm S}, \bar{\phi}_{\rm S}$ are the angles we simulate and the subscript "S" denotes "source". $\bar{\phi}(t)=\bar{\phi}_{0}+2\pi t/T$, where T equals one year, and $\bar{\phi}_{0}$ is just a constant specifying the detector's location at time $t=0$. $\alpha_{1}(t)=2\pi t/T-\pi/12+\alpha_{0}$, where $\alpha_{0}$ is just a constant that specifies the orientation of the arms at $t=0$. In our simulation, we adopt $\bar{\phi}_{0}=\alpha_{0}=0$ for simplicity. The instant $t(f)$ that quadrupole frequency sweeps past $f$ until the instant $t_{c}$ when the BHs merge is (to the lowest order):
\begin{align}
t(f)=t_{c}-5(8\pi f)^{-8/3}\mathcal{M}_{c}^{-5/3},
\end{align}
where $f$ and $t$ are the frequency and time measured on the Earth. In the following simulation on GW events detected by LISA, we shall use Eqs. \eqref{eq:thetaS} and \eqref{eq:phiS} as the angles inserting into the pattern functions.

\section{Simulations of the gravitational wave detections}\label{sec:simulations}

The NS mass distribution we choose is uniform within $[1,2]\,M_{\odot}$, where $M_{\odot}$ is the solar mass. The black hole mass is chosen to be uniform in the interval $[3,10]\,M_{\odot}$. The ratio of possibly detecting black hole-neutron star binary (BHNS) and BNS events we consider here is nearly 0.03 \cite{Abadie:2010px}. The redshift distribution of the observable sources is given by \cite{Zhao:2010sz}
\begin{align}\label{eq:redshift}
P(z)\propto \frac{4\pi d_{C}^{2}(z)R(z)}{(1+z)E(z)},
\end{align}
where $d_{C}$ is the comoving distance defined as $d_{C}(z)\equiv\int_{0}^{z}1/E(z^{\prime})dz^{\prime}$, and $R(z)$ describes the NS-NS merger rate, which is given by \cite{Schneider:2000sg}
\begin{align}\label{eq:mergerrate}
R(z)=\left\{
\begin{array}{lcl}
1+2z,	        &	& {z\leq1};\\
\frac34(5-z),	&	& {1<z<5};\\
0,	            &	& {z\geq5}.
\end{array}\right.
\end{align}
Speaking of standard siren sources for LISA, three representative models of the expected MBHB sources are adopted following \cite{Barausse:2012fy,Tamanini:2016zlh,Tamanini:2016uin}:
\begin{enumerate}
	\item \emph{Model pop III.}--- a "realistic" light-seed model including the delays with which massive BHs merge after their host galaxies coalesce, stating that massive BHs form from the remnants of population III (pop III) stars;
	\item \emph{Model Q3d.}--- a "realistic" heavy-seed model with delays included, which states that massive BHs form from the collapse of protogalactic disks;
	\item \emph{Model Q3nod.}---the same as model Q3d, but with no delays.
\end{enumerate}
The mass distribution of massive BHs we choose is uniform within [$10^{4}, 10^{7}$] $M_{\odot}$, and the redshift distributions of MBHB events during five-year observations for the above three models are adopted from Fig. 1 in \cite{Tamanini:2016uin}, of which the total number of GW events is 28, 27 and 41 for the above three models, respectively.

A GW event is claimed only when the signal-to-noise ratio (SNR) of the detector network reaches over 8, following the current threshold used in LIGO/Virgo analysis. The combined SNR for the network of $N$ ($N=2$ for LISA, $N=3$ for ET and $N=2$ for DECIGO) independent interferometers is given by
\begin{align}\label{eq:SNR}
\rho=\sqrt{\sum_{i=1}^{N}(\rho^{(i)})^{2}},
\end{align}
where $\rho^{(i)}=\sqrt{\langle\mathcal{H}^{(i)},\mathcal{H}^{(i)}\rangle}$. Given $\tilde{a}(f)$ and $\tilde{b}(f)$ as the Fourier transforms of some functions $a(t)$ and $b(t)$, the scalar product is defined as
\begin{align}\label{eq:innerproduct}
\langle a,b\rangle\equiv4\int_{f_{\rm min}}^{f_{\rm max}}\frac{\tilde{a}(f)\tilde{b}^{*}(f)+\tilde{a}^{*}(f)\tilde{b}(f)}{2}\frac{df}{S_{h}(f)},
\end{align}
where $S_{h}(f)$ denotes the one-side noise power spectral density (PSD), characterizing the performance of GW detector.

For LISA, the noise PSD is given by \cite{Klein:2015hvg}
\begin{align}\label{eq:LISAPSD}
S_{h}^{\rm LISA}(f)&=\frac{20}{3}\frac{4S_{\rm n,acc}(f)+S_{\rm n,sn}(f)+S_{\rm n,omn}(f)}{L^{2}}\notag\\
&\times\bigg[1+\bigg(\frac{f}{0.41\frac{c}{2L}}\bigg)^{2}\bigg]\ \rm Hz^{-1},
\end{align}
where $L$ is the arm length taken to be $2\rm\,Gm$, $S_{\rm n,acc}(f), S_{\rm n,sn}(f)$  and $S_{\rm n,omn}(f)$ represent the noise components due to low-frequency acceleration, shot noise and other measurement noise, respectively. We adopt the following values for N2A2M5L6 configurations
\begin{align}
&S_{\rm n,acc}(f)=\frac{9\times10^{-30}}{(2\pi f)^{4}}\bigg(1+\frac{10^{-4}}{f}\bigg)\,\rm m^{2}\,Hz^{{-1}};\notag\\
&S_{\rm n,sn}(f)=2.22\times10^{-23}\,\rm m^{2}\,Hz^{{-1}};\notag\\
&S_{\rm n,omn}(f)=2.65\times10^{-23}\,\rm m^{2}\,Hz^{{-1}}.
\end{align}
The noise PSD of ET is \cite{Zhao:2010sz}
\begin{align}\label{eq:ETPSD}
S_{h}^{\rm ET}(f)=&10^{-50}(2.39\times10^{-27}x^{-15.64}+0.349x^{-2.145}\notag\\
&+1.76x^{-0.12}+0.409x^{1.1})^{2}\,\rm Hz^{-1},
\end{align}
where $x=f/f_{p}$ with $f_{p}\equiv100\rm\,Hz$. For DECIGO, the noise PSD is \cite{Yagi:2009zz}
\begin{align}\label{eq:DECIGOPSD}
S_{h}^{\rm DECIGO}(f)=&5.3\times10^{-48}\bigg[(1+x^{2})+\frac{2.3\times10^{-7}}{x^{4}(1+x^{2})}\notag\\
&+\frac{2.6\times10^{-8}}{x^{4}}\bigg]\,\rm Hz^{-1},
\end{align}
where $x=f/f_{p}$ with $f_{p}\equiv7.36\rm\,Hz$. The lower and upper cutoff frequencies for LISA are chosen to be $f_{\rm min}=10^{-4}\,\rm Hz$ and $f_{\rm max}=c/2\pi L\simeq0.05\frac{\rm Gm}{L}\,\rm Hz$, respectively. For ET, we adopt them to be $f_{\rm min}=1\,\rm Hz$ and $f_{\rm max}=2f_{\rm LSO}$, where the orbit frequency at the last stable orbit $f_{\rm LSO}=1/6^{3/2}2\pi\mathcal{M}_{\rm obs}$ with the observed total mass $M_{\rm obs}=(1+z)M$. As for DECIGO, they are taken to be $f_{\rm min}=0.233(\frac{M_{\odot}}{\mathcal{M}_{c}})^{5/8}(\frac{\rm yr}{T_{\rm obs}})^{3/8}\,\rm Hz$ and $f_{\rm max}=100\,\rm Hz$, respectively \cite{Nishizawa:2010xx}, here $T_{\rm obs}$ denotes the observation time, and we set it as one year in the following simulation.

We apply the standard Fisher matrix to estimate the instrumental error on the measurement of luminosity distance. We assume that the error on $d_{\rm L}$ is uncorrelated with errors on the remaining GW parameters, so that
\begin{align}\label{eq:instError1}
\sigma_{d_{\rm L}}^{\rm inst}\simeq\sqrt{\bigg\langle\frac{\partial\mathcal{H}}{\partial d_{\rm L}},\frac{\partial\mathcal{H}}{\partial d_{\rm L}}\bigg\rangle^{-1}}.
\end{align}
From Eqs. \eqref{eq:FTWaveform} and \eqref{eq:Amplitude}, it can be seen that $\mathcal{H}\propto d_{\rm L}^{-1}$, hence $\sigma_{d_{\rm L}}^{\rm inst}\simeq d_{\rm L}/\rho$. As mentioned above, the inclination angle is ideally set to be 0. However, when we estimate the practical uncertainty of the measurement of luminosity distance, the correction between $d_{\rm L}$ and $\iota$ is then necessary to be taken into account. Note that the maximal effect of the inclination on the SNR is a factor of 2 when we take $\iota$ between $0^\circ$ and $90^\circ$ \cite{Li:2013lza}. Therefore, we add this factor to the instrumental error for a conservative estimation
\begin{align}\label{eq:instError2}
\sigma_{d_{\rm L}}^{\rm inst}\simeq\frac{2d_{\rm L}}{\rho}.
\end{align}
Another error that we need to consider is $\sigma_{d_{\rm L}}^{\rm lens}$ due to the effect of weak lensing, and we assume $\sigma_{d_{\rm L}}^{\rm lens}/d_{\rm L}=0.05z$ as \cite{Zhao:2010sz}.

For ET, we therefore take the total uncertainty on the luminosity distance as
\begin{align}\label{eq:ETerrordL}
\sigma_{d_{\rm L}}&=\sqrt{(\sigma_{d_{\rm L}}^{\rm inst})^{2}+(\sigma_{d_{\rm L}}^{\rm lens})^{2}}; \notag\\
&=\sqrt{\bigg(\frac{2d_{\rm L}}{\rho}\bigg)^{2}+(0.05zd_{\rm L})^{2}}.
\end{align}
As for DECIGO, we adopt the lensing error by following \cite{Nishizawa:2010xx} as
\begin{align}\label{eq:DECIGOlensingError}
\sigma_{d_{\rm L}}^{\rm lens}(z)=d_{\rm L}(z)\times 0.066\bigg[\frac{1-(1+z)^{-0.25}}{0.25}\bigg]^{1.8}.
\end{align}
In addition, peculiar velocity error due to the clustering of galaxies and binary barycentric motion is considered as well, and is given by \cite{Gordon:2007zw}
\begin{align}
\sigma_{d_{\rm L}}^{\rm pv}(z)=d_{\rm L}(z)\times\bigg|1-\frac{(1+z)^{2}}{H(z)d_{\rm L}(z)}\bigg|\sigma_{\rm v,gal},
\end{align}
where $\sigma_{\rm v,gal}$ is the one-dimensional velocity dispersion of the galaxy and set to be $\sigma_{\rm v,gal}=300\rm\,km\,s^{-1}$, independent of the redshifts. Therefore, the total uncertainty on the measurement of $d_{\rm L}$ is
\begin{align}\label{eq:DECIGOerrordL}
\sigma_{d_{\rm L}}=\sqrt{(\sigma_{d_{\rm L}}^{\rm inst})^{2}+(\sigma_{d_{\rm L}}^{\rm lens})^{2}+(\sigma_{d_{\rm L}}^{\rm pv})^{2}} .
\end{align}
In the case of LISA \cite{Tamanini:2016zlh}, the main contribution of the total errors on $d_\mathrm{L}$ at high redshift comes from the weak lensing part \eqref{eq:DECIGOlensingError}, which will decrease by a factor of 2 when we account for the merger and ringdown. The peculiar velocity error \cite{Tamanini:2016zlh} is given by
\begin{align}
\sigma_{d_{\rm L}}^{\rm pv}(z)=d_{\rm L}(z)\times\bigg[1+\frac{c(1+z)}{H(z)d_{\rm L}(z)}\bigg]\frac{\sqrt{\langle v^{2}\rangle}}{c},
\end{align}
where $\sqrt{\langle v^{2}\rangle}$ is the peculiar velocity of the host galaxy with respect to the Hubble flow, which we fix at $500\,\mathrm{km\,s^{-1}}$ as a rough estimate. In principle, we only need to consider the lensing error that dominates the most part of the total errors. However, as a conservative estimation, we also include the instrumental error part, which almost makes no difference on the final results.

In order to find out the constraint ability on anisotropy, we need to calculate $\chi^{2}$, which is given by
\begin{align}\label{eq:chisquare}
\chi^{2}=\sum_{i=1}^{N}\bigg[\frac{d_{\rm L}^{i}-d_{\rm L}^{\rm fid}(\hat{z})}{\sigma_{d_{\rm L}}^{i}}\bigg]^{2},
\end{align}
where $d_{\rm L}^{i},\ \sigma_{d_{\rm L}}^{i}$ are the $i$th luminosity distance and corresponding error of the simulated data, $N$ denotes the number of data sets. The fiducial luminosity distance $d_{\rm L}^{\rm fid}(\hat{z})$ is given by
\begin{align}\label{eq:dLfid}
d_{\rm L}^{\rm fid}(\hat{z})=d_{\rm L}^{0}(z)[1+g(\hat{n}\cdot\hat{z})],
\end{align}
where we parametrize the dipole modulation simply by its amplitude $g$ and direction $\hat{n}$ given by
\begin{align}\label{eq:n}
\hat{n}=(\cos\phi\sin\theta,\ \sin\phi\sin\theta,\ \cos\theta),
\end{align}
where $\theta\in[0,\pi)$ and $\phi\in[0,2\pi)$. The fiducial angles of anisotropic direction we choose are $\theta^{f}=\pi/2$ and $\phi^{f}=\pi$. Since we are just interested in the constraint ability on their measurements, the exact values are not essential in the simulations. Therefore, the fiducial amplitude $g^f$ is chosen as a variable ranging from 0.005 to 0.1 with interval 0.005.

The approach we adopted in this paper to simulate the mock data of standard siren events is as follows \cite{Cai:2016sby}:
\begin{enumerate}
  \item We first simulate GW events of number $N$, of which the redshift values is according to the redshift distribution in Eq.~\eqref{eq:redshift}, and the angles $\theta$ and $\phi$ are randomly sampled within the intervals [0,$\pi$] and [0,$2\pi$]. For each GW event with redshift $z$, we then calculate the isotropic luminosity distance $d_\mathrm{L}^0(z)$ according to Eq. \eqref{eq:dL0} and anisotropic fiducial luminosity distance $d_\mathrm{L}^\mathrm{fid}(\hat{z})$ according to Eq. \eqref{eq:dLfid}.
  \item We next use the fiducial luminosity distance $d_\mathrm{L}^\mathrm{fid}(\hat{z})$ to evaluate the SNR and corresponding error $\sigma_{d_\mathrm{L}}$ with random values for the masses $m_1$ and $m_2$ of coalescing binaries. The random value for the mass of neutron star is between $[1,2]$ $M_{\odot}$, and the random mass of black hole is within $[3,10]$ $M_{\odot}$. It is worth noting that, the ratio of possibly detecting BHNS and BNS events is set to be $0.03$ \cite{Abadie:2010px}. It is also worth noting that, the random value for the black hole mass of MBHB is between $[10^{4}, 10^{7}]$ $M_{\odot}$ for LISA.
  \item Since there is a threshold for a successful claim of GW event detection, we have to redo the random sampling for $m_1$ and $m_2$ until $\mathrm{SNR}>8$ is fulfilled. Finally we simulate the measurement of luminosity distance $d_{\rm L}^{\rm mea}=\mathcal{N}(d_{\rm L}^{\rm fid}, \sigma_{d_{\rm L}})$ from the fiducial value of $d_{\rm L}^{\rm fid}$ and the error $\sigma_{d_{\rm L}}$. It is worth noting that, for some direction angles, the threshold $\mathrm{SNR}>8$ may not be fulfilled no matter how many trials of random sampling for the binaries masses are done. Therefore we have to simulate GW events of number $N$ larger than we need at the first place, say $2N$, and pick the first $N$ events after the whole simulation process is finished.
  \item With the simulated measurements of both luminosity distance and redshift in hand, we then calculate the $\chi^{2}$ in Eq. \eqref{eq:chisquare} with $d_{\rm L}^{i}$ recognized as $d_{\rm L}^{\rm mea}$. We apply the MCMC method to calculate the likelihood function of ($g,\theta,\phi$) and find out the constrained dipole modulation $(g^c,\theta^c,\phi^c)$, which will be compared with the presumed fiducial dipole modulation $(g^f,\theta^f,\phi^f)$.
\end{enumerate}

\begin{figure*}
	\includegraphics[width=0.32\textwidth]{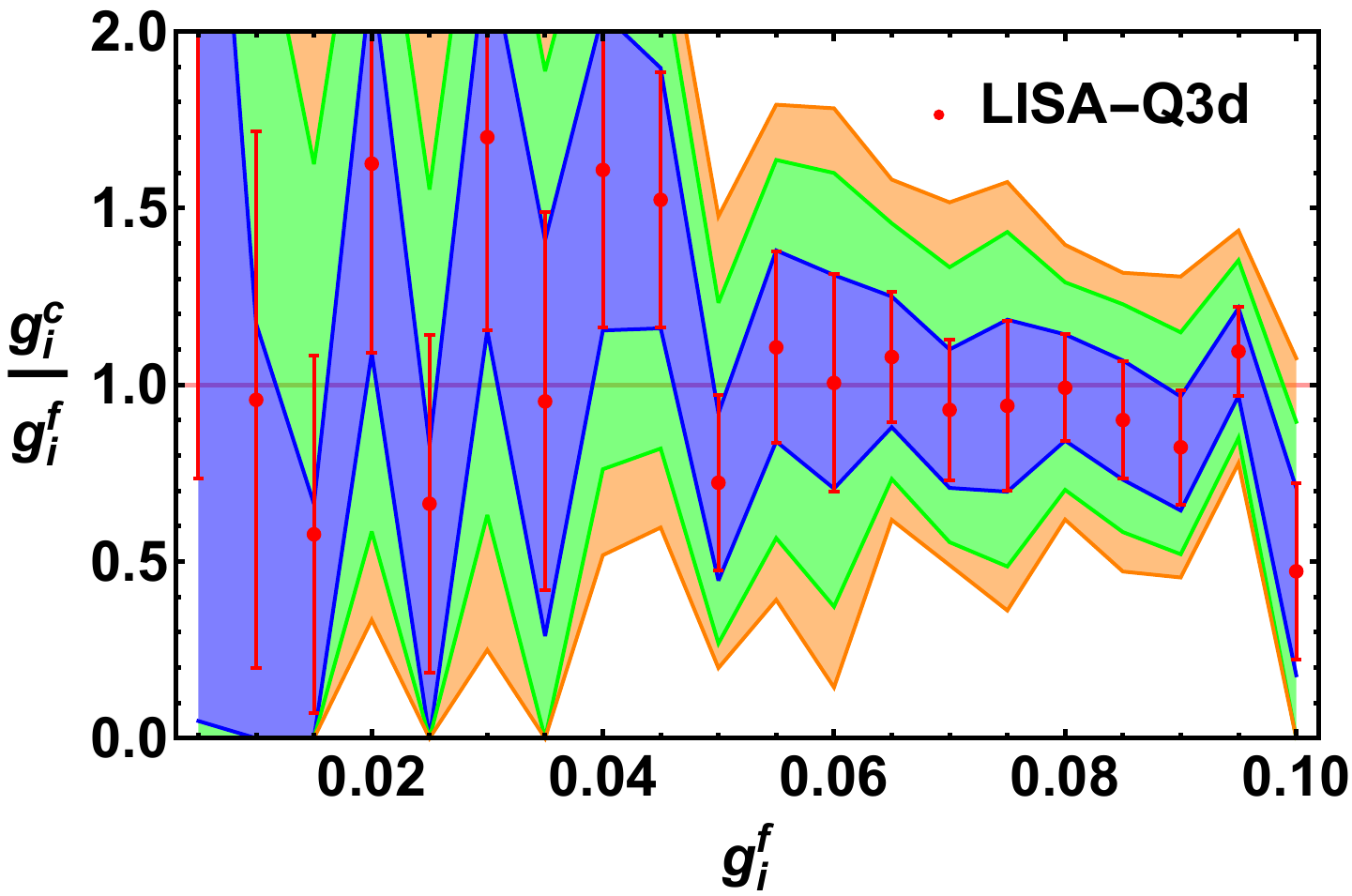}
	\includegraphics[width=0.32\textwidth]{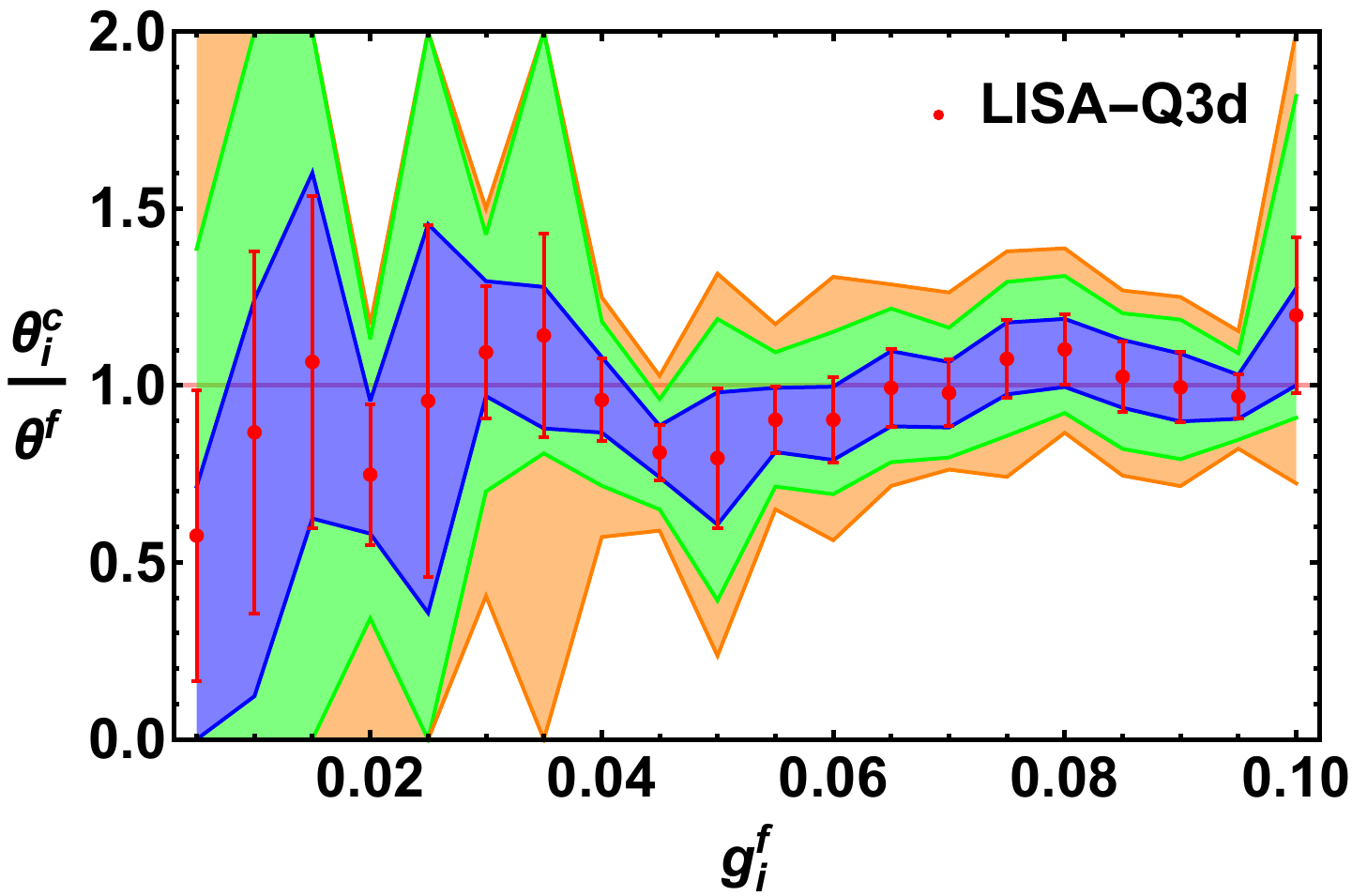}
	\includegraphics[width=0.32\textwidth]{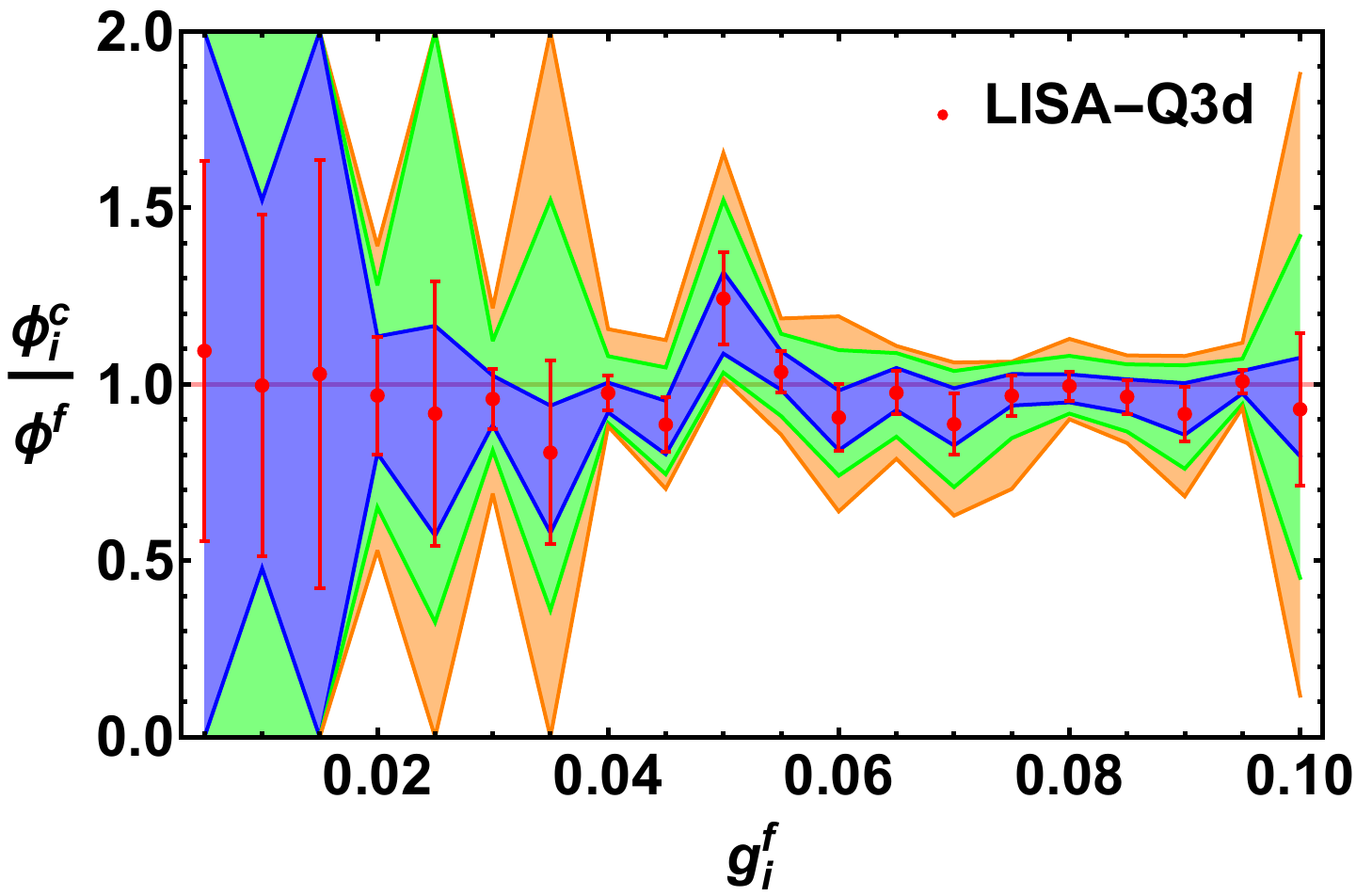}\\
	\includegraphics[width=0.32\textwidth]{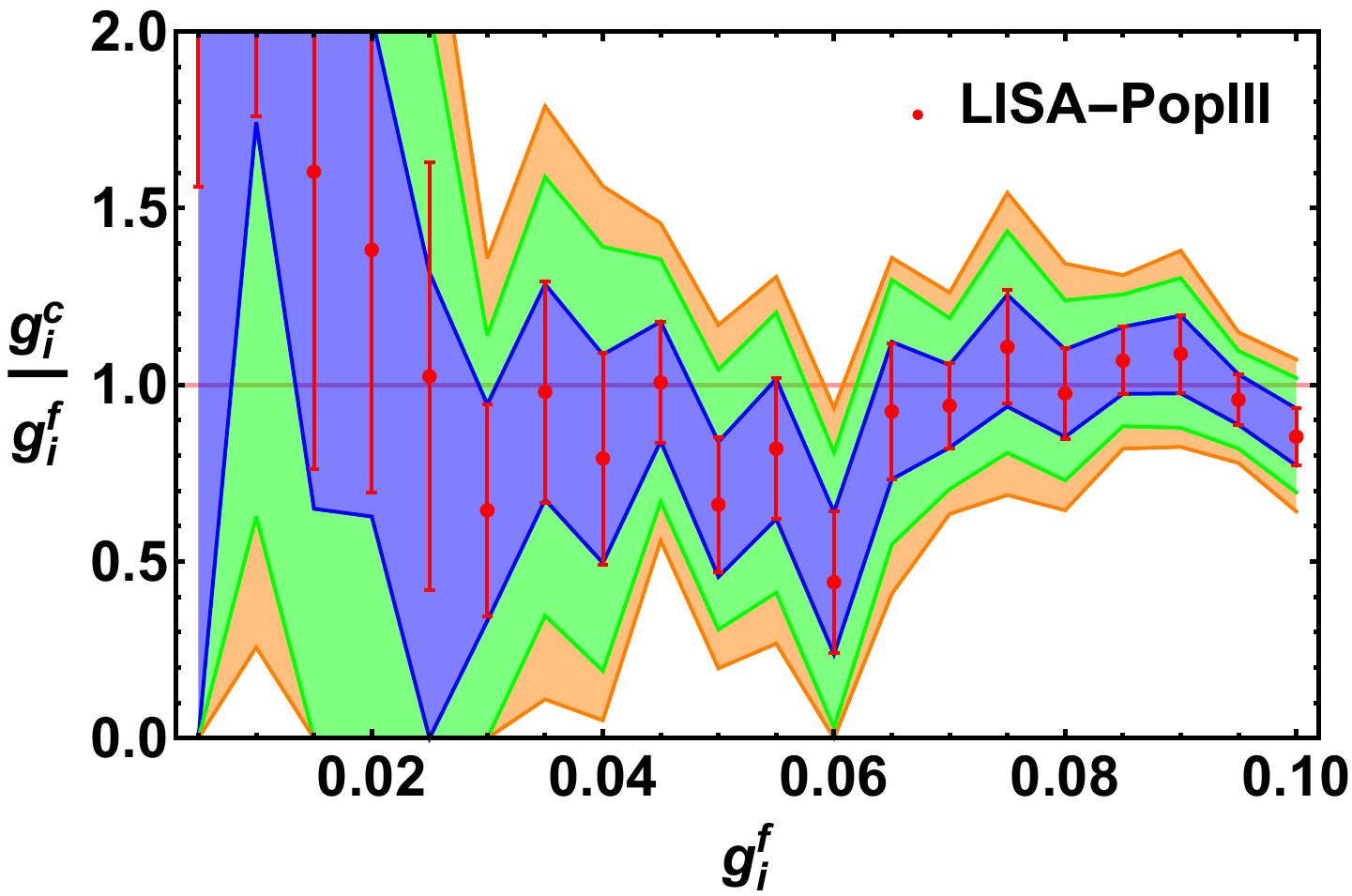}
	\includegraphics[width=0.32\textwidth]{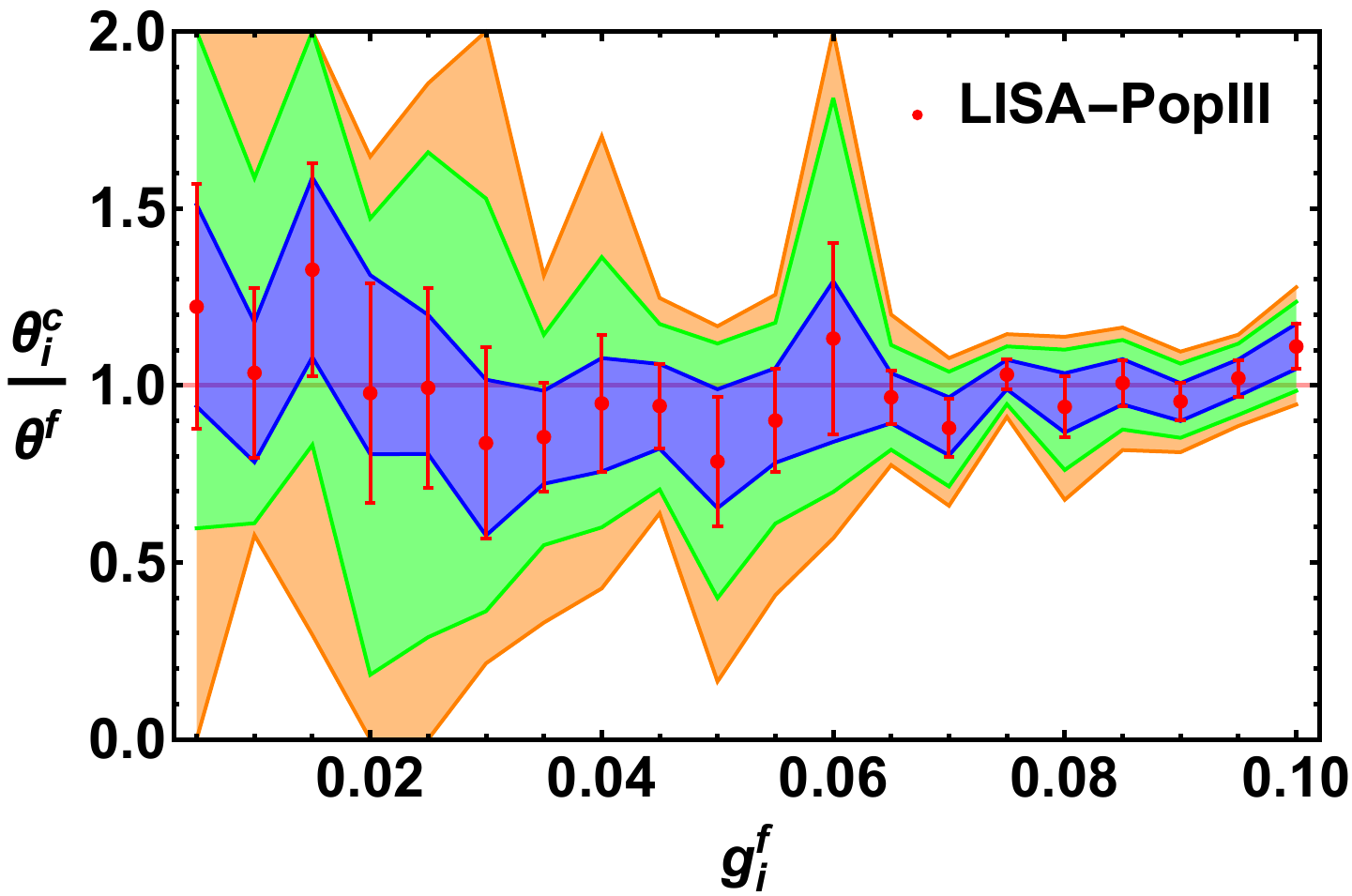}
	\includegraphics[width=0.32\textwidth]{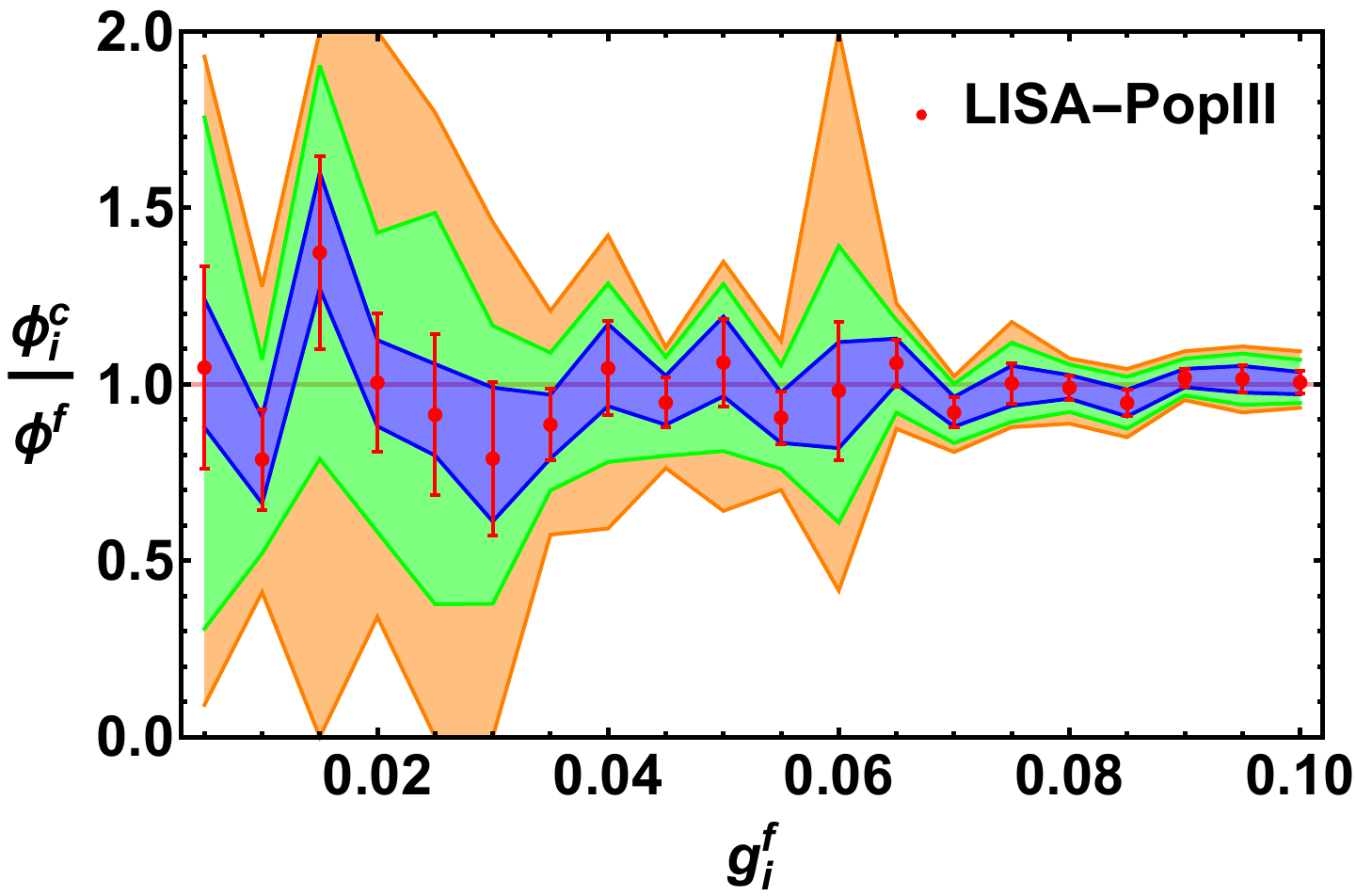}\\
	\includegraphics[width=0.32\textwidth]{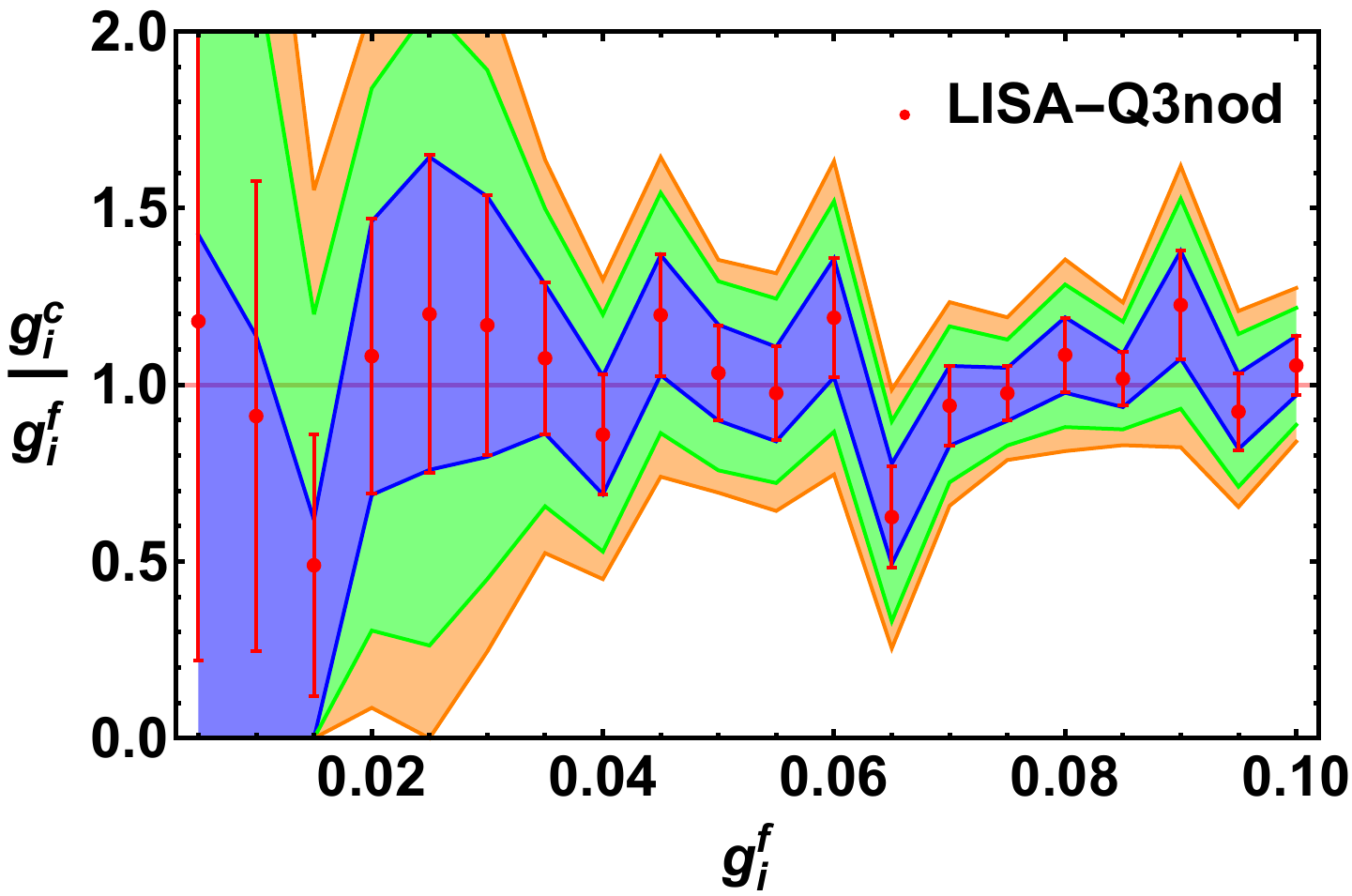}
	\includegraphics[width=0.32\textwidth]{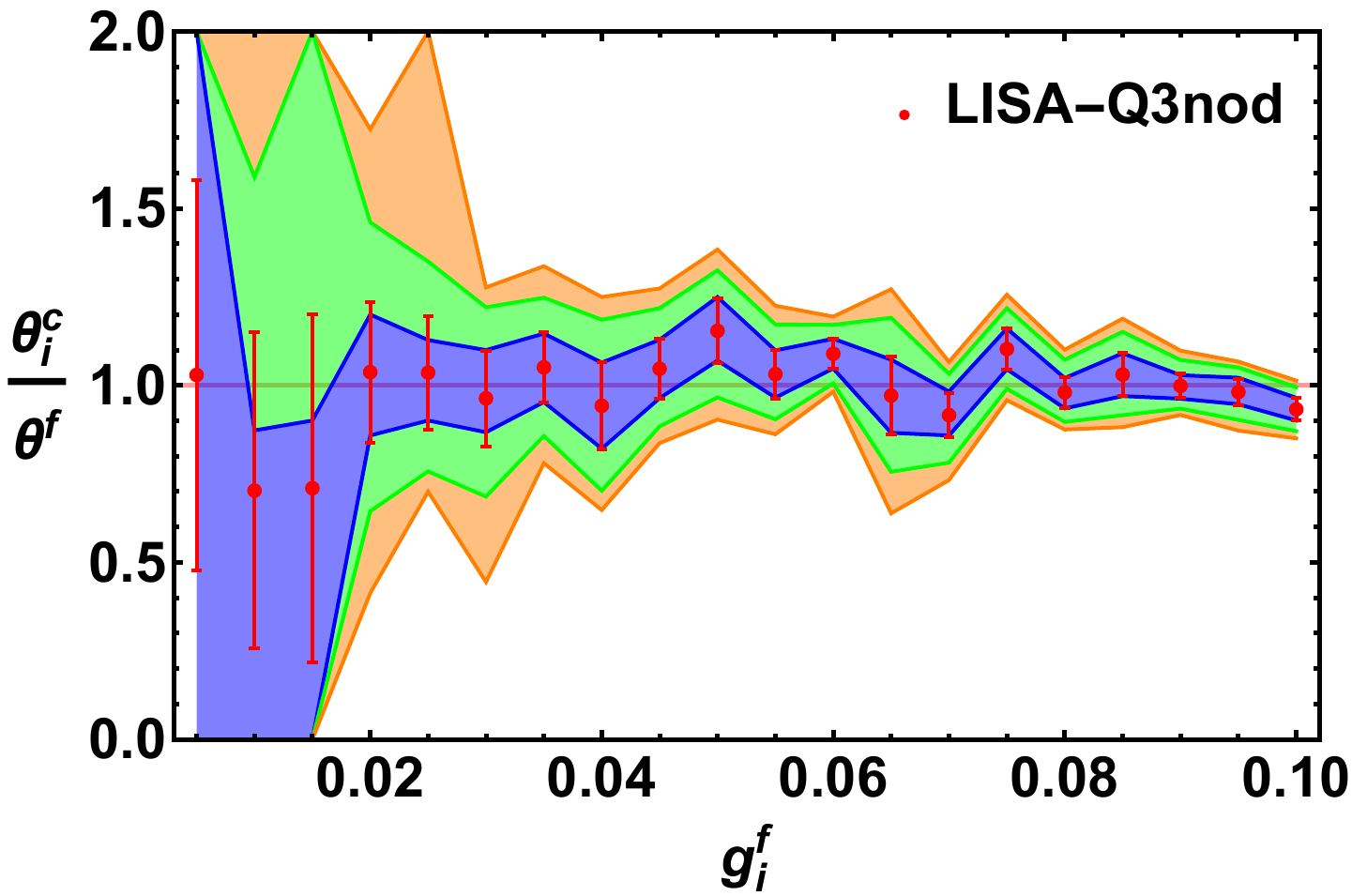}
	\includegraphics[width=0.32\textwidth]{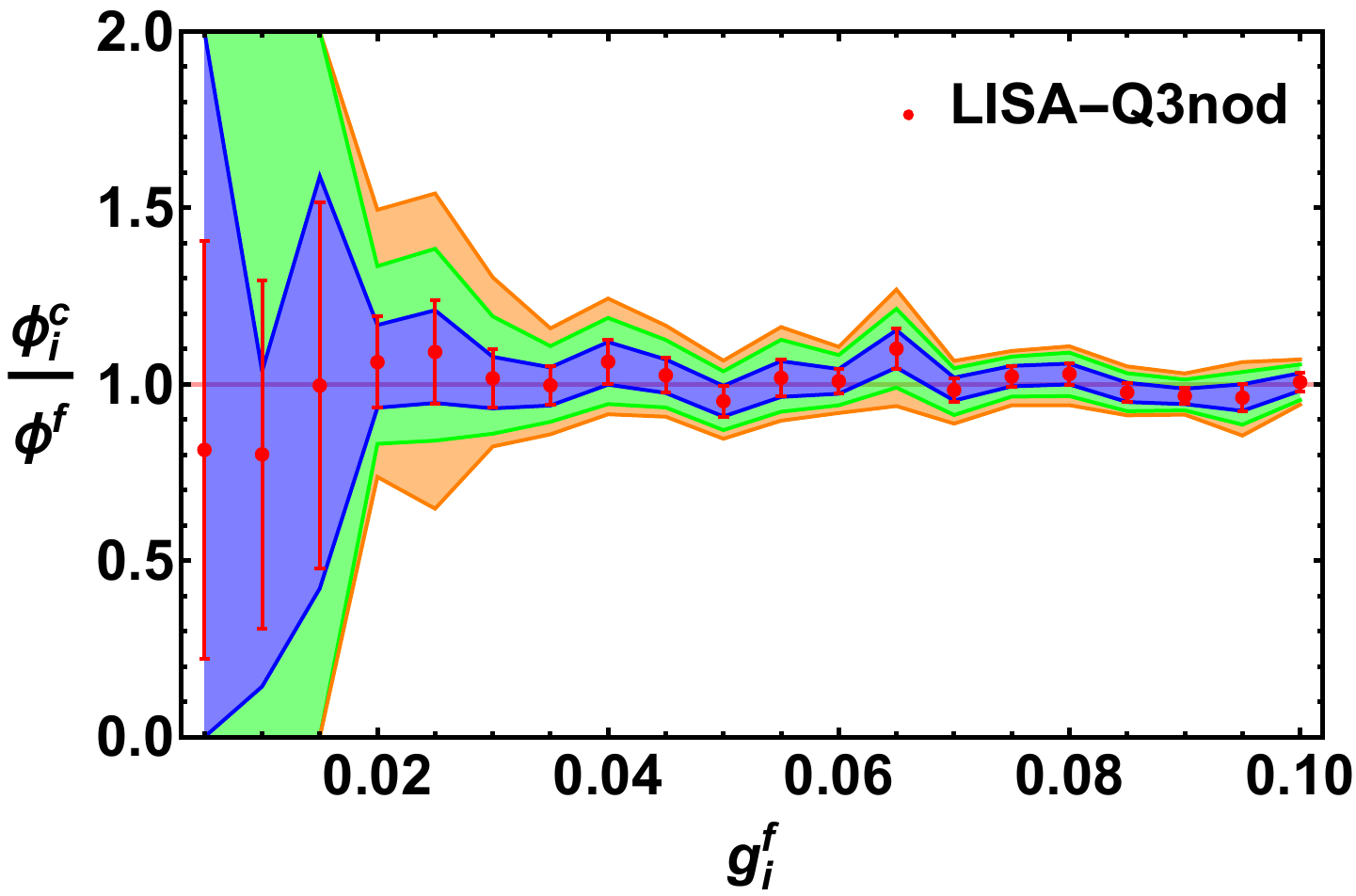}\\
	\caption{The constraint ability of LISA for dipole anisotropy from standard siren events with respect to the varying fiducial amplitude $g^f$ of dipole modulation. The standard siren events are simulated from the MBHB models Q3d (top panels), pop III (medium panels) and Q3nod (bottom panels). The best constrained values divided by the corresponding fiducial values for $g$ (left panels), $\theta$ (medium panels) and $\phi$ (right panels) are labeled by the red dots with standard deviation error bars. The blue/green/orange shaded regions are of 1$\sigma$/2$\sigma$/3$\sigma$ C.L., respectively.}\label{fig:LISA}
\end{figure*}

\begin{figure*}
  \includegraphics[width=0.32\textwidth]{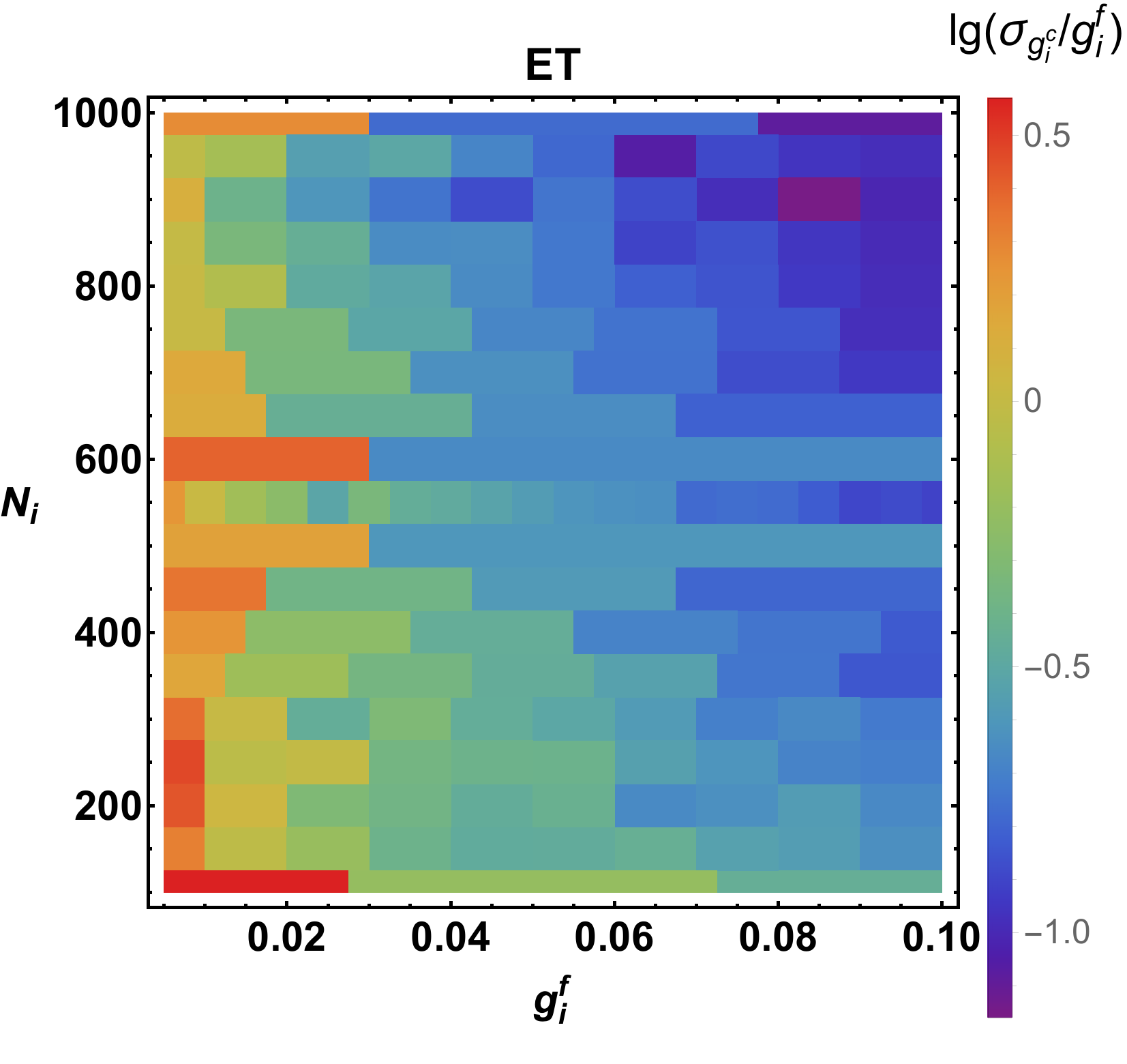}
  \includegraphics[width=0.32\textwidth]{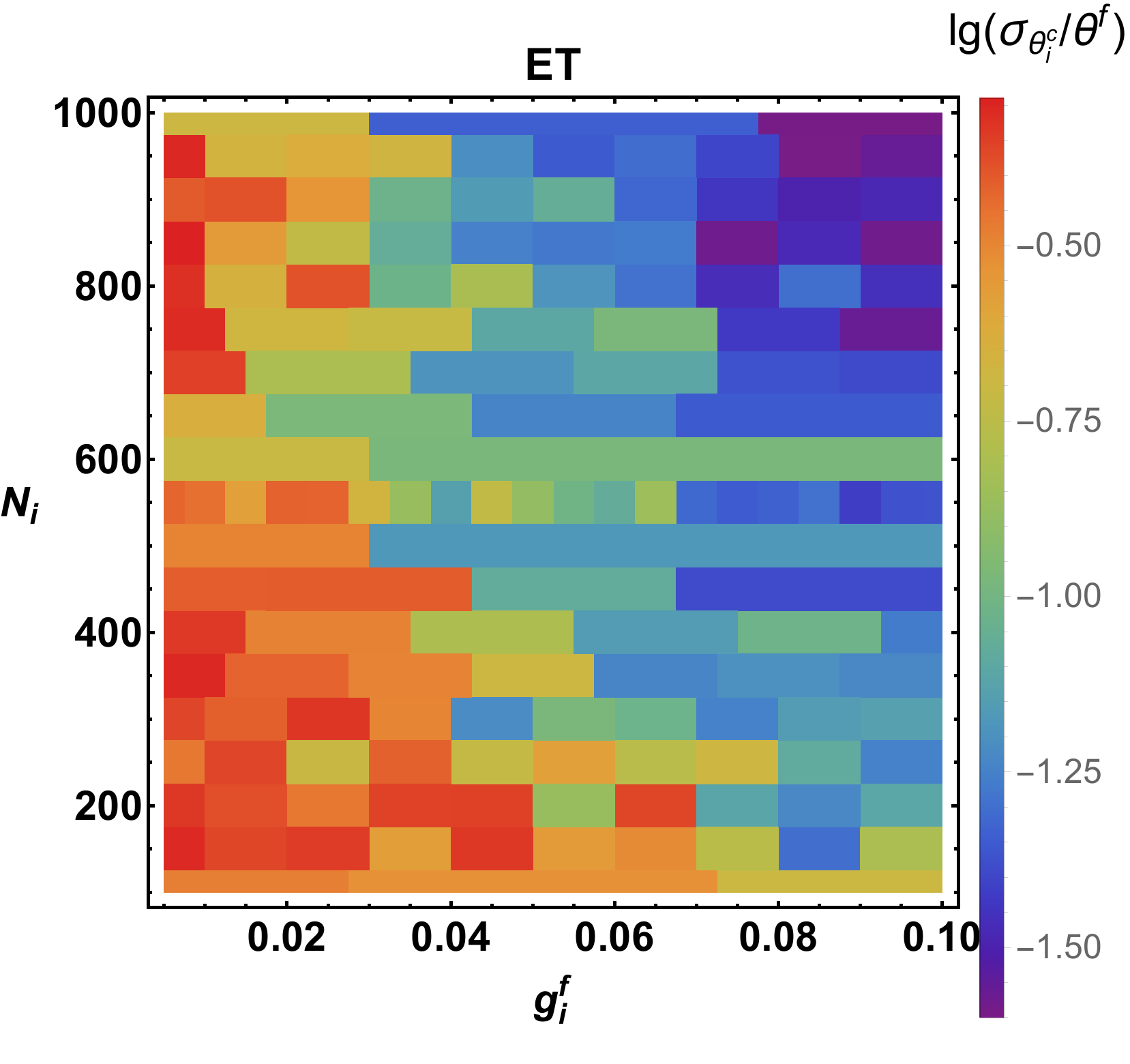}
  \includegraphics[width=0.32\textwidth]{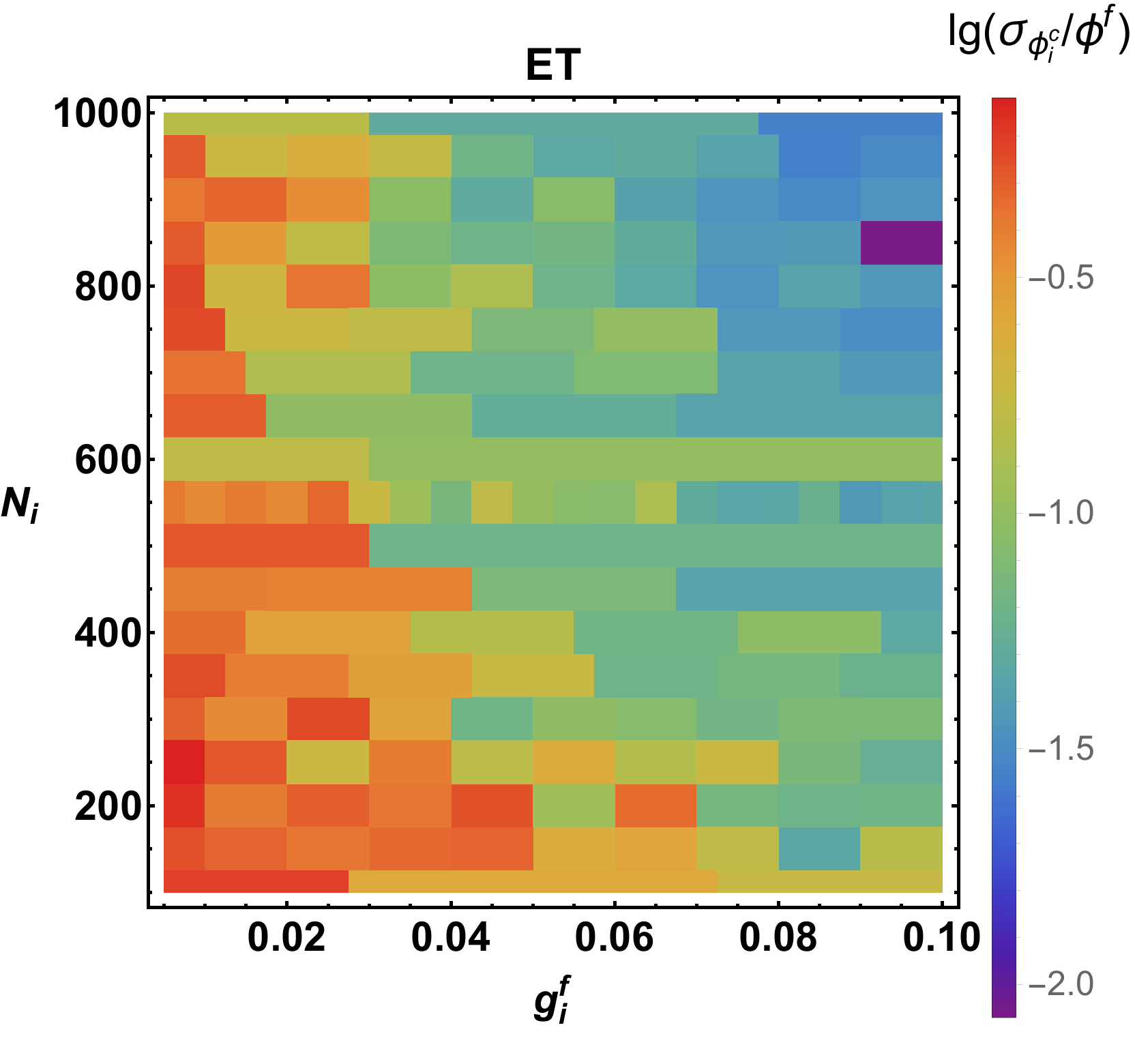}\\
  \includegraphics[width=0.32\textwidth]{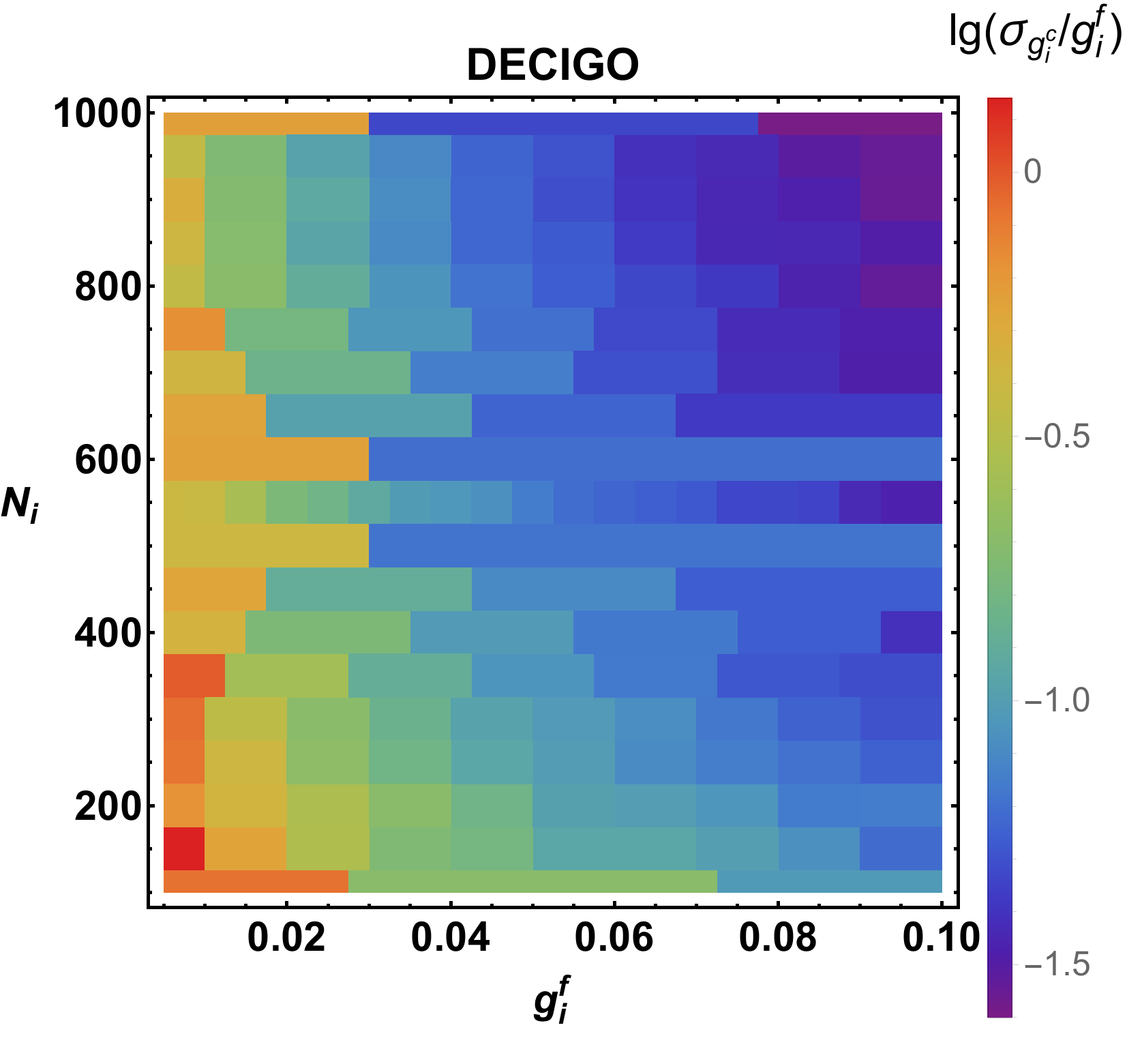}
  \includegraphics[width=0.32\textwidth]{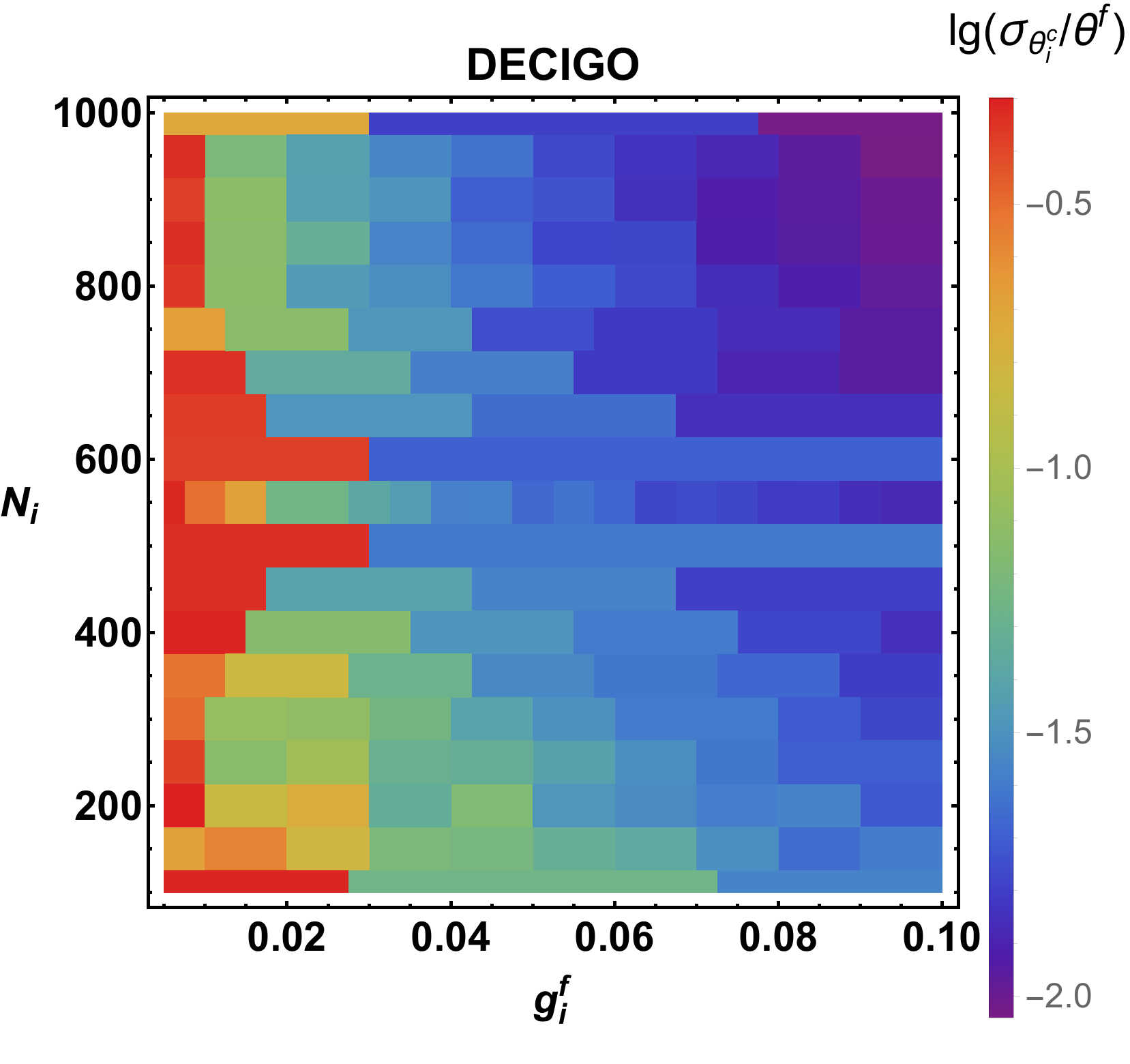}
  \includegraphics[width=0.32\textwidth]{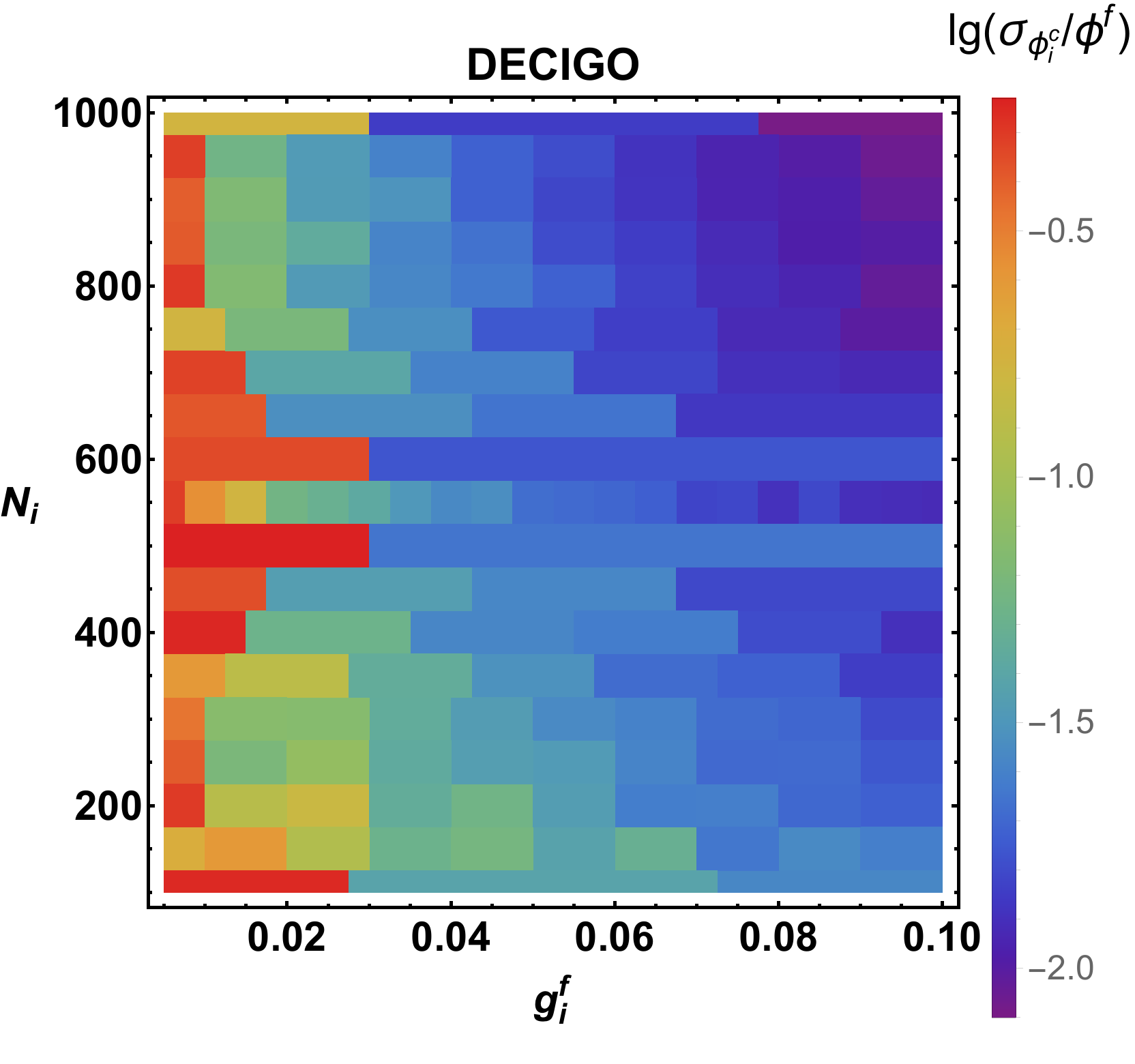}\\
  \caption{The standard deviation for $g$ (left panels), $\theta$ (medium panels) and $\phi$ (right panels) as a function of fiducial dipole amplitude $g^{f}$ and GW event number $N$, normalized by fiducial values of $g^{f}$, $\theta^f$ and $\phi^f$ in logarithmic units, respectively. Results for ET (DECIGO) are shown in the top (bottom) panels.}\label{fig:ETandDECIGO}
\end{figure*}

\begin{figure*}
  \includegraphics[width=0.32\textwidth]{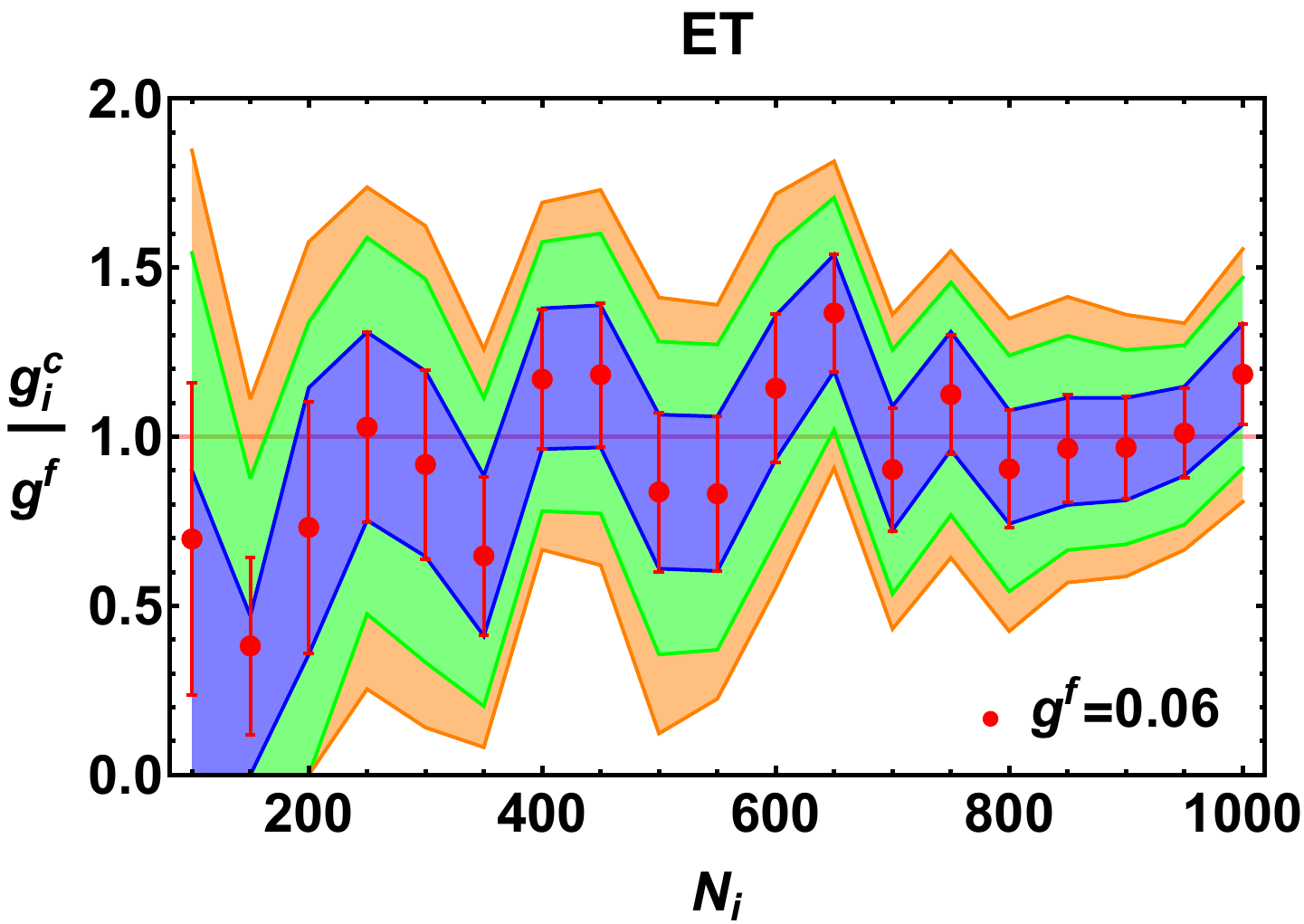}
  \includegraphics[width=0.32\textwidth]{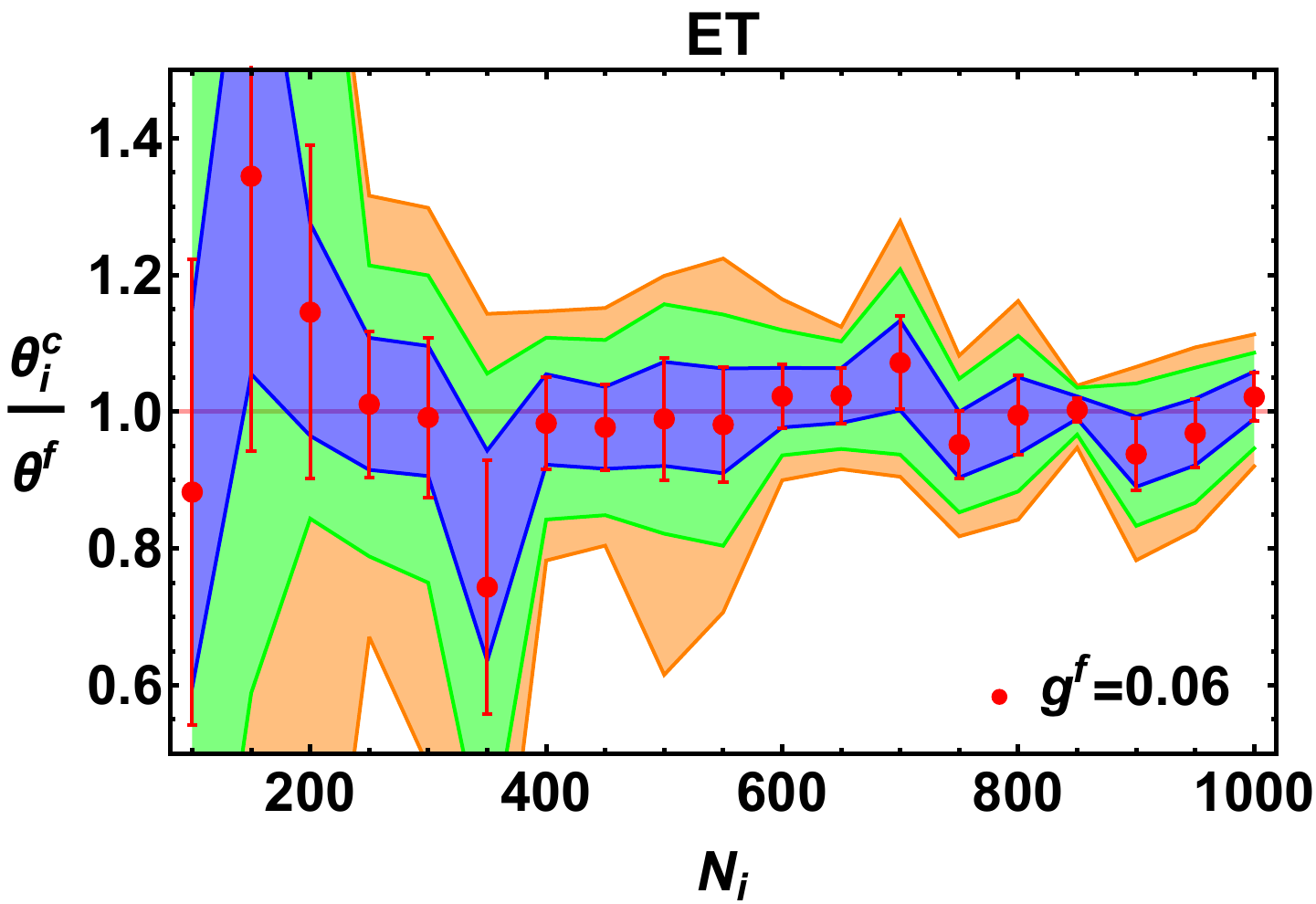}
  \includegraphics[width=0.32\textwidth]{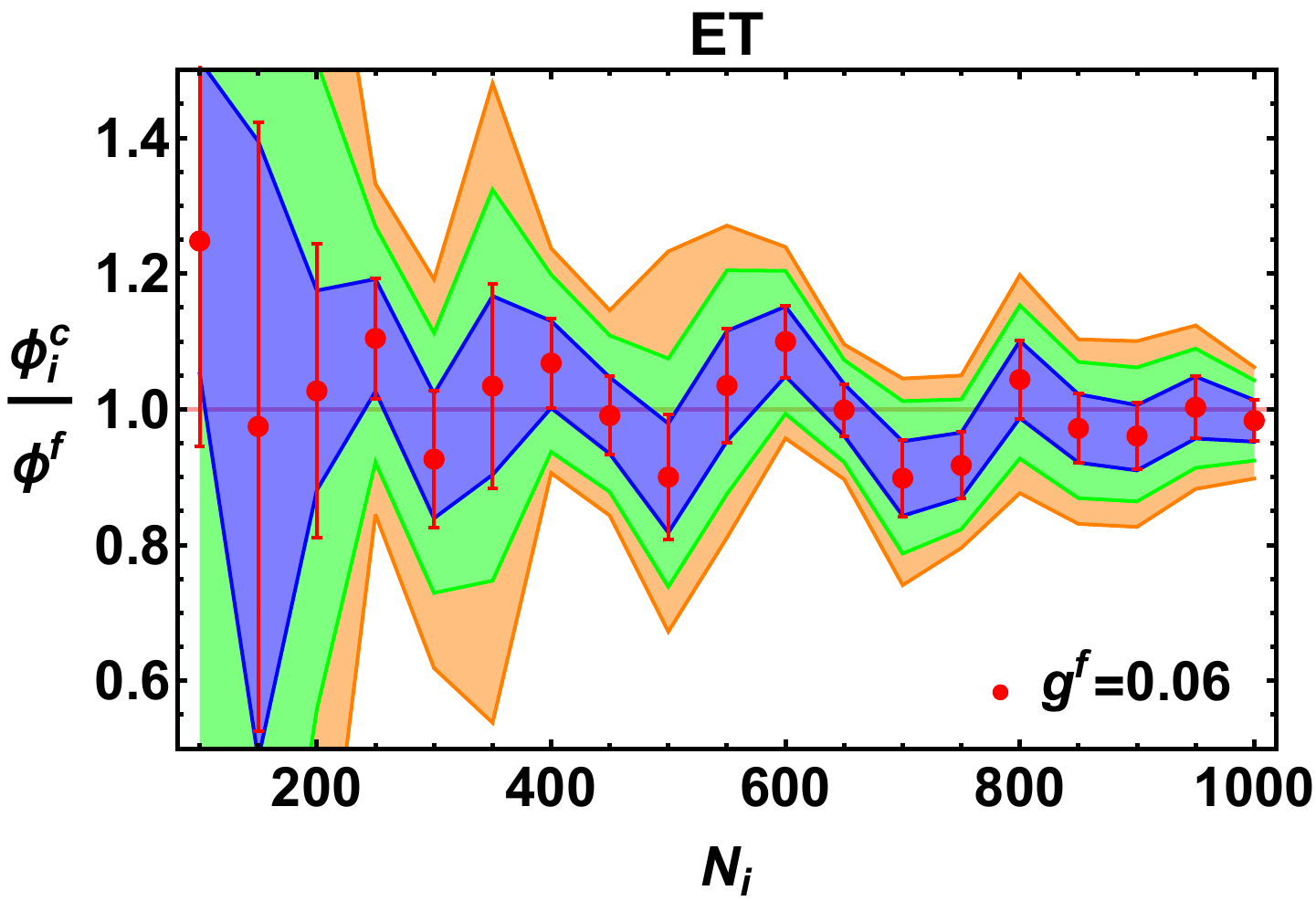}\\
  \includegraphics[width=0.32\textwidth]{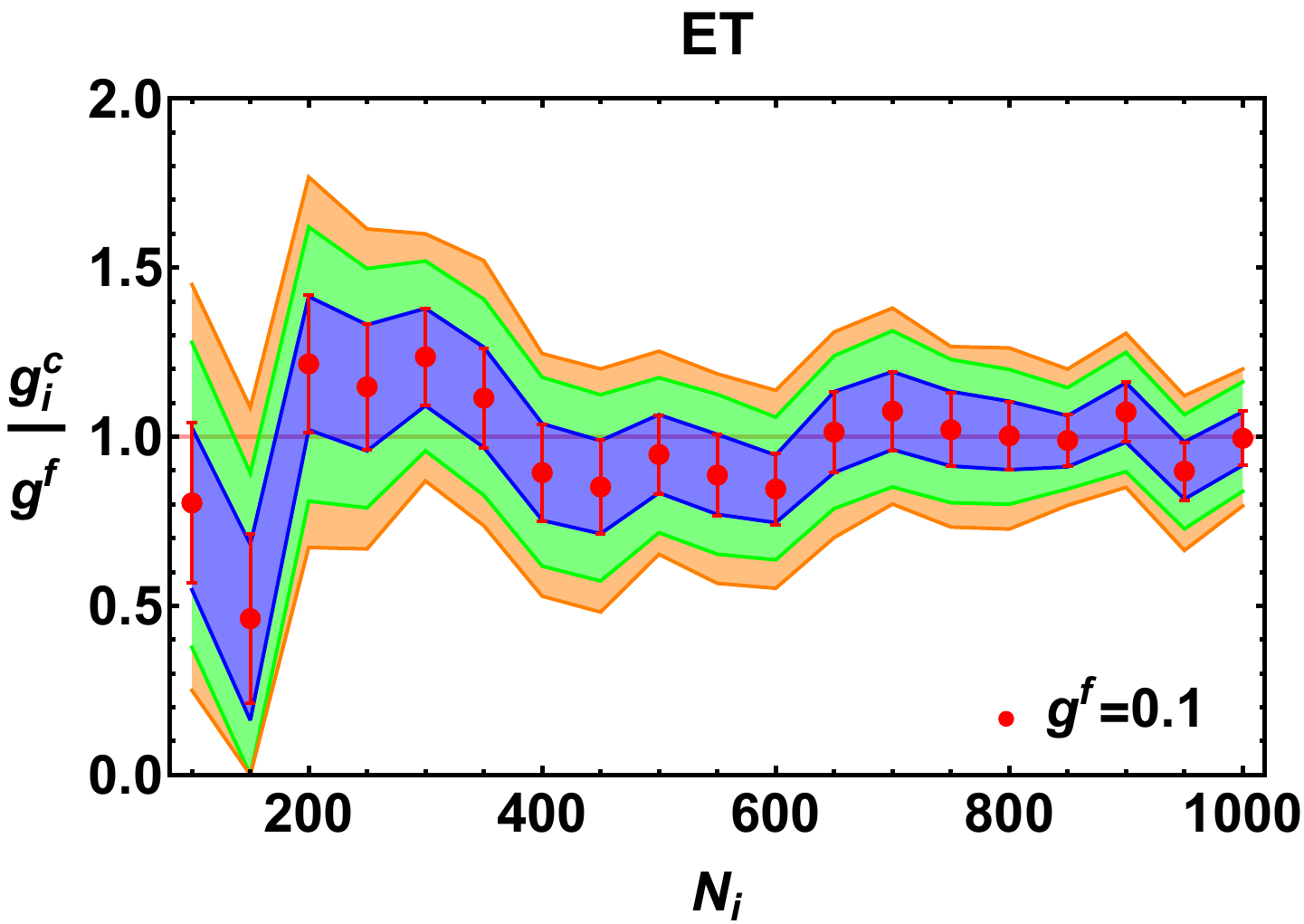}
  \includegraphics[width=0.32\textwidth]{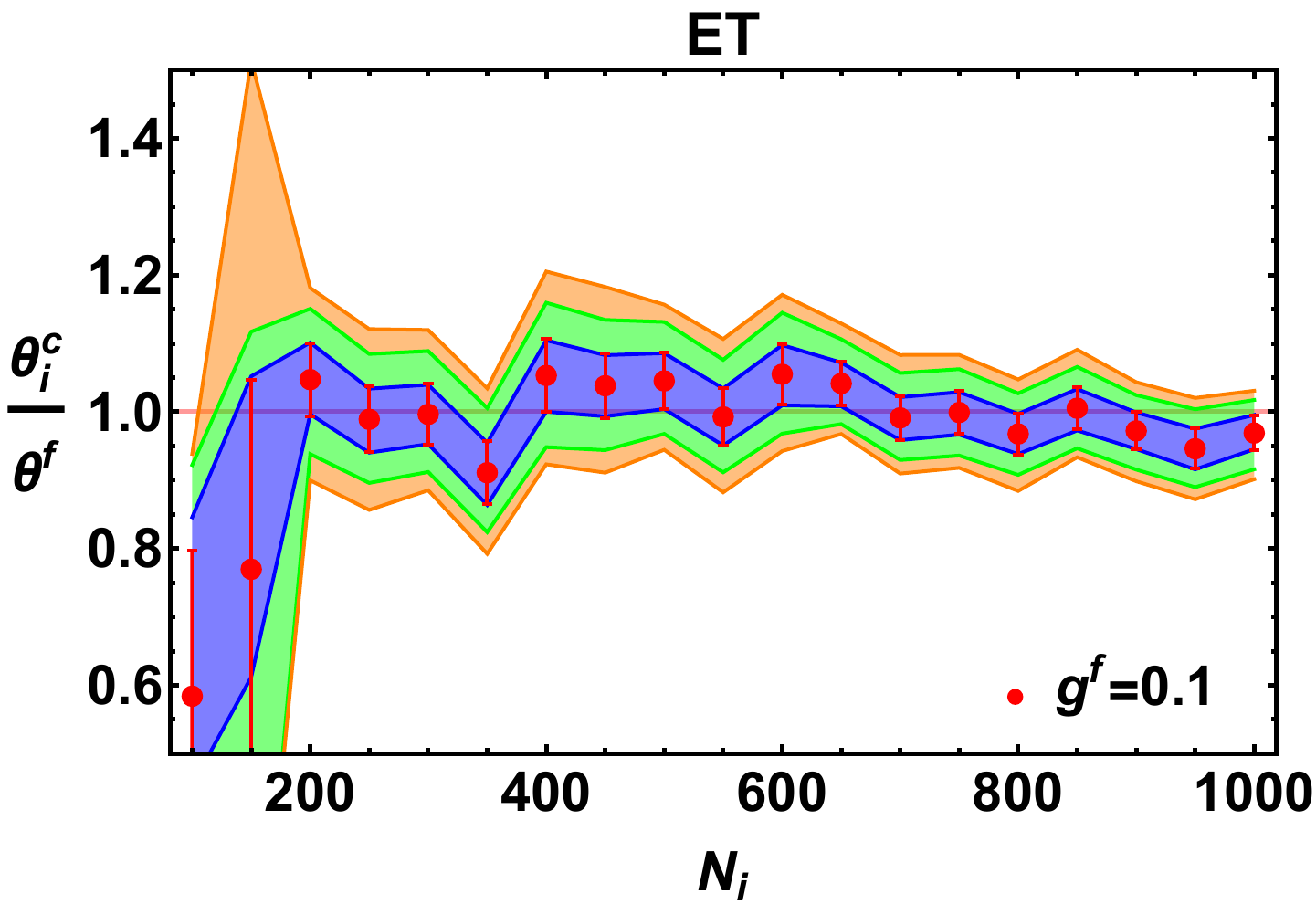}
  \includegraphics[width=0.32\textwidth]{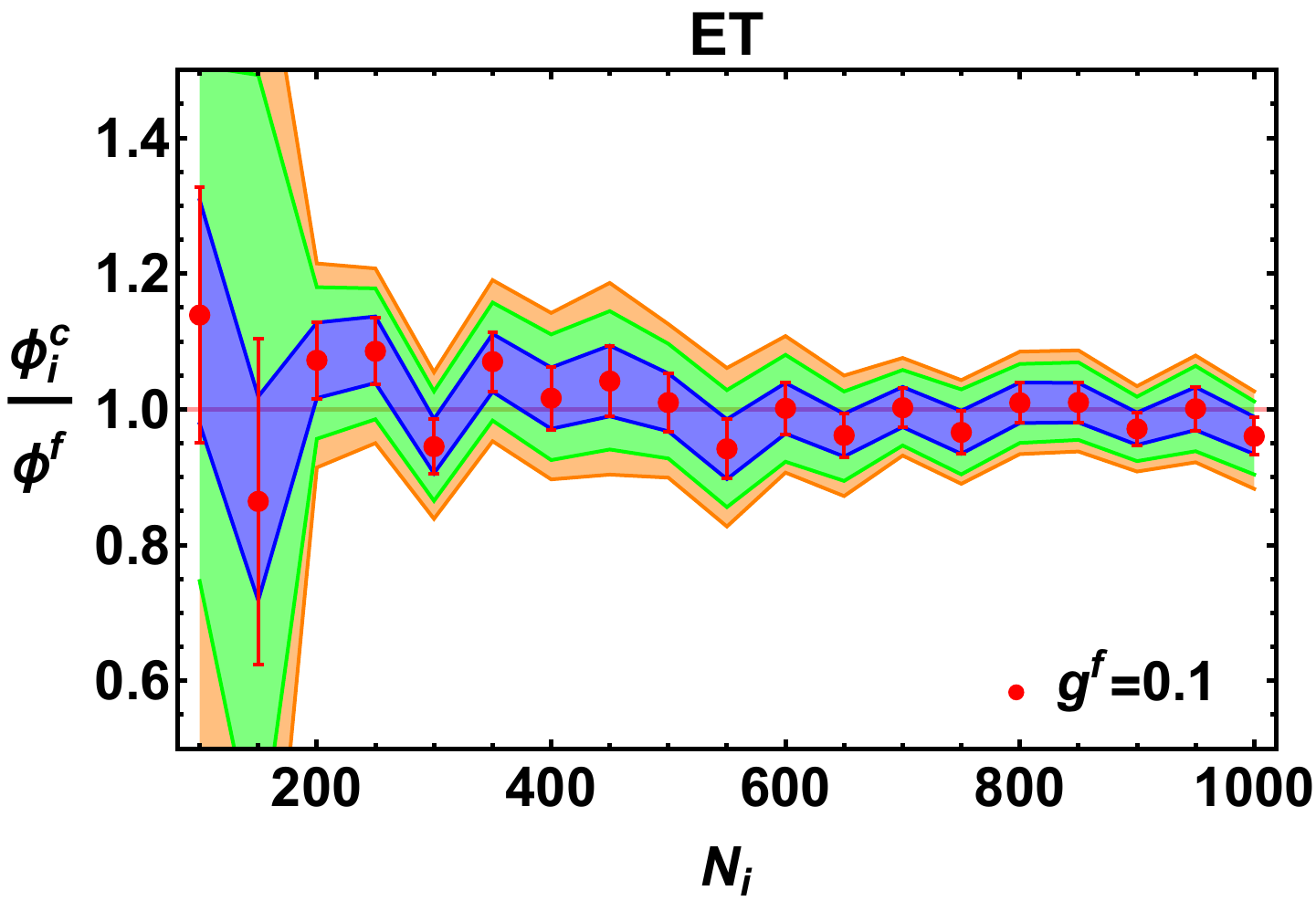}\\
  \caption{The constraint ability of ET with respect to the varying number of standard siren events for given fiducial values of $g^f=0.06$ (top panels) and $g^f=0.1$ (bottom panels). The best constrained values divided by the corresponding fiducial values for $g$ (left panels), $\theta$ (medium panels) and $\phi$ (right panels) are labeled by the red dots with standard deviation error bars. The blue/green/orange shaded regions are of 1$\sigma$/2$\sigma$/3$\sigma$ C.L., respectively.}\label{fig:ETSigma}
\end{figure*}

\begin{figure*}
  \includegraphics[width=0.32\textwidth]{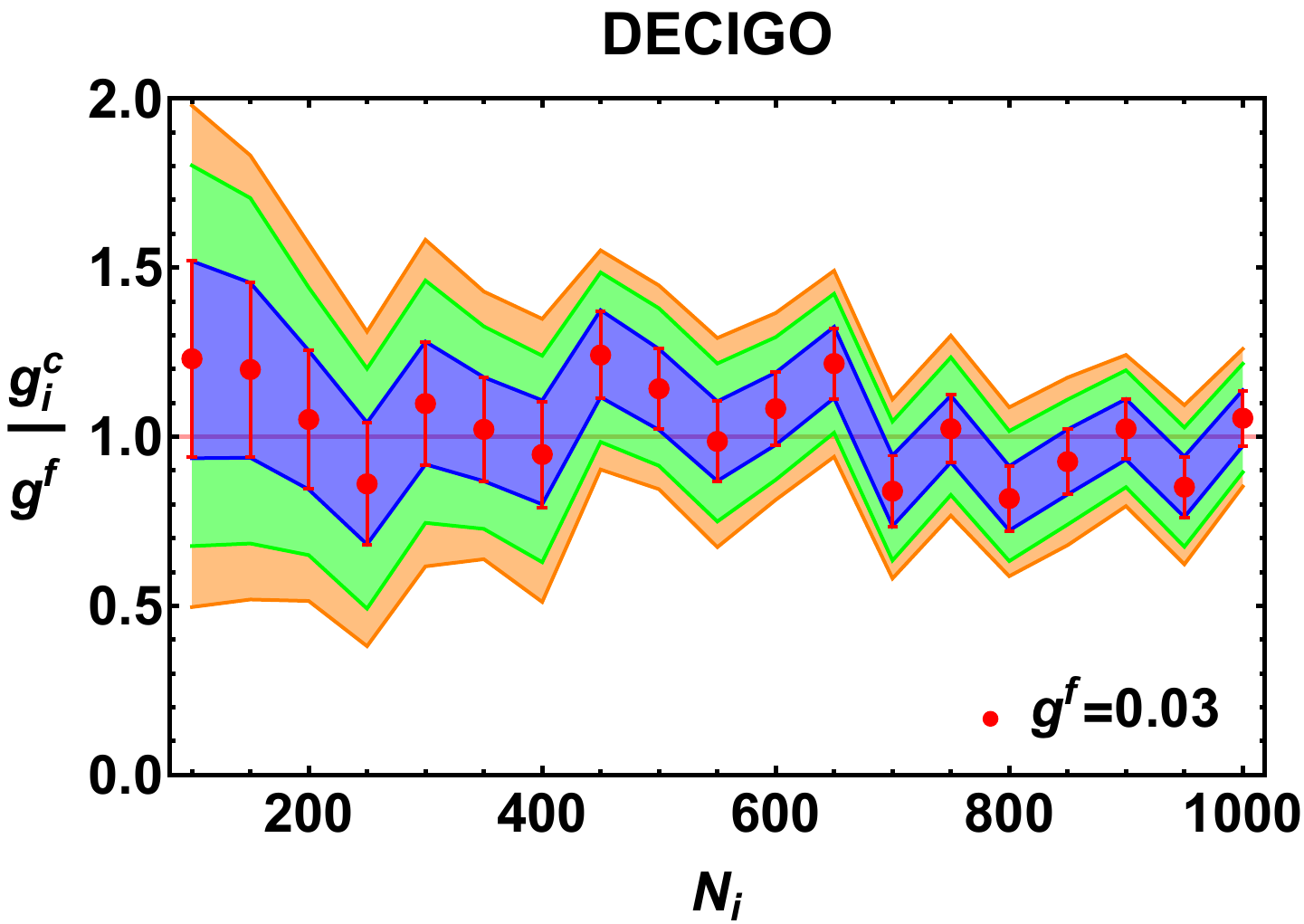}
  \includegraphics[width=0.32\textwidth]{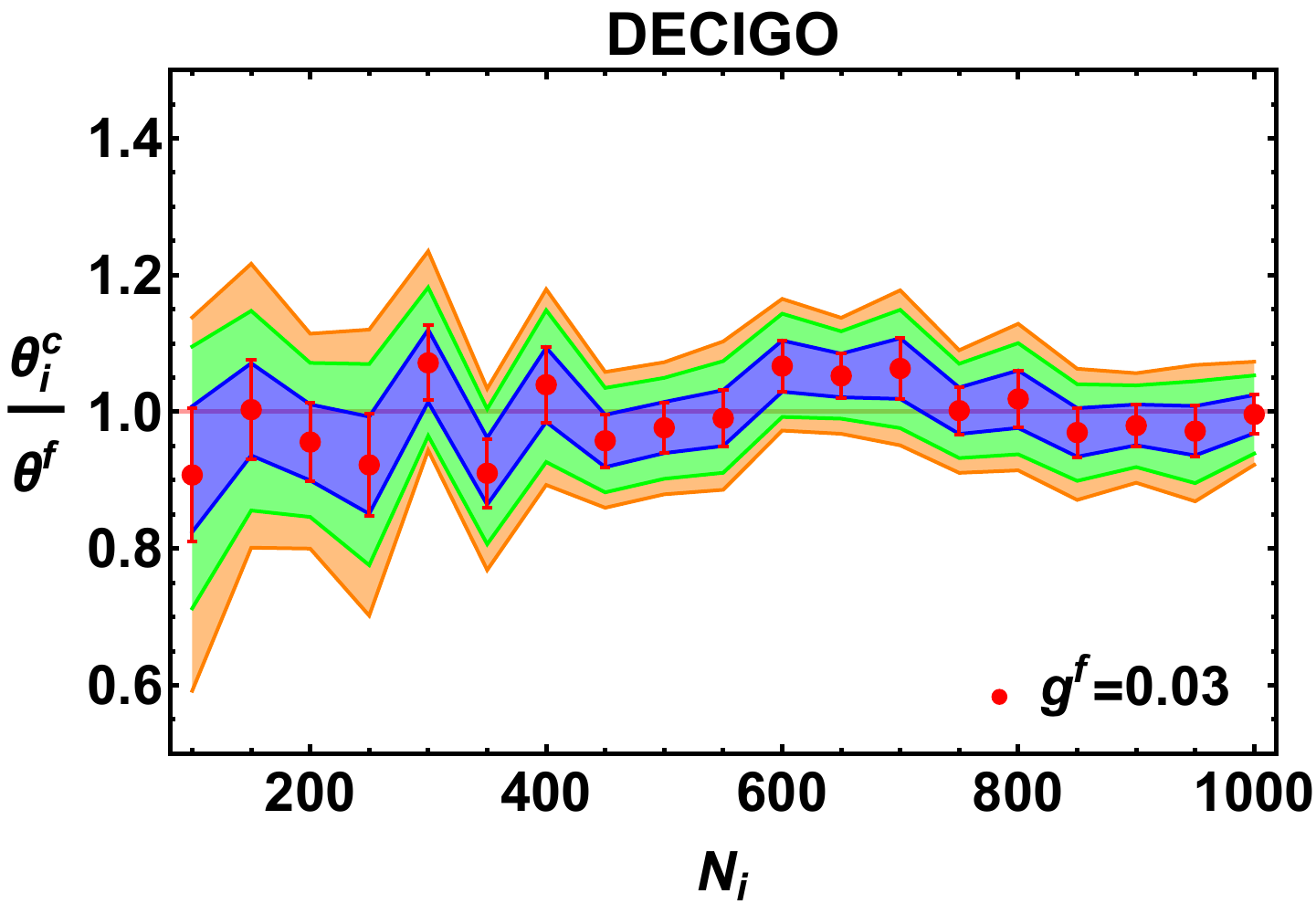}
  \includegraphics[width=0.32\textwidth]{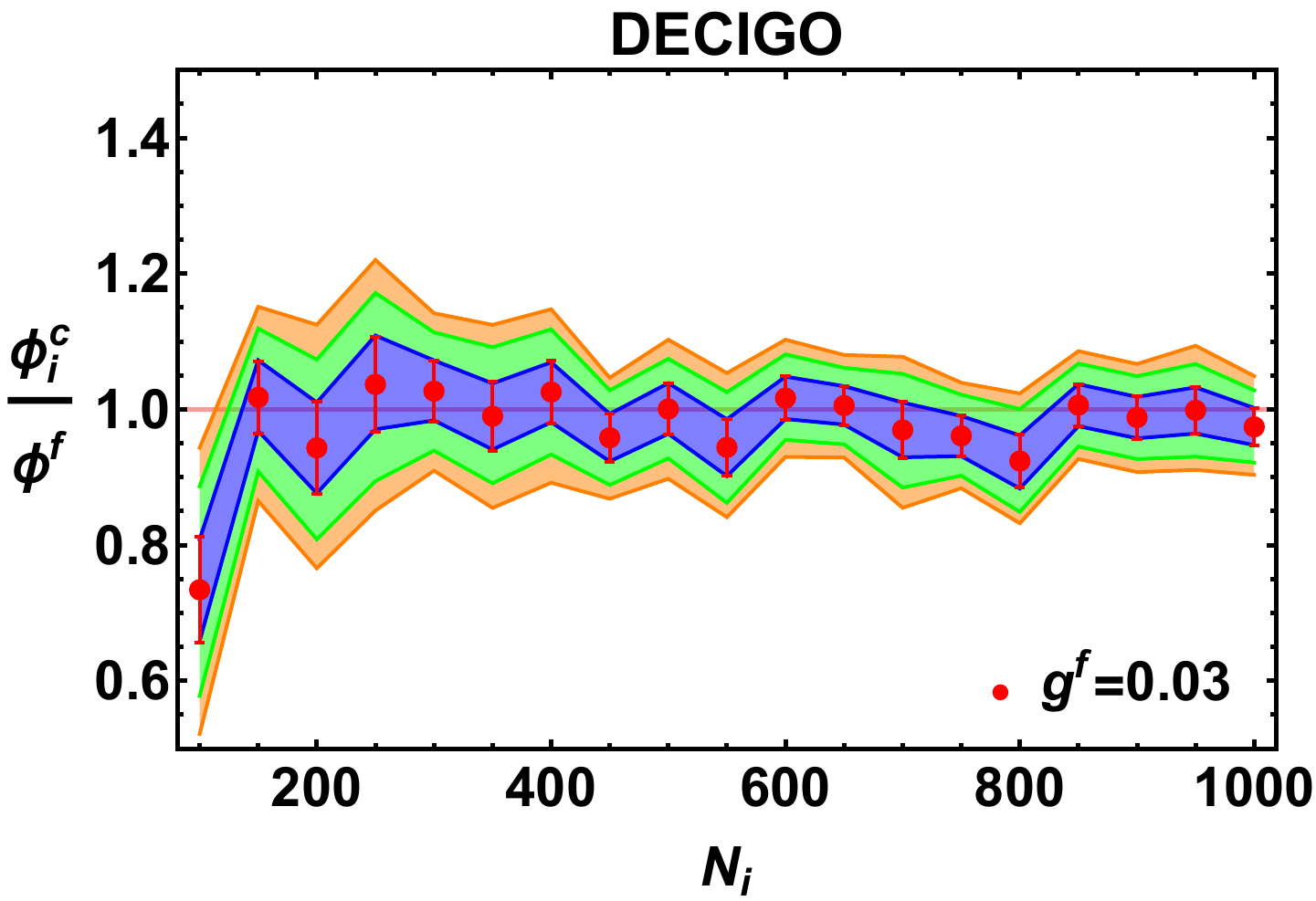}\\
  \includegraphics[width=0.32\textwidth]{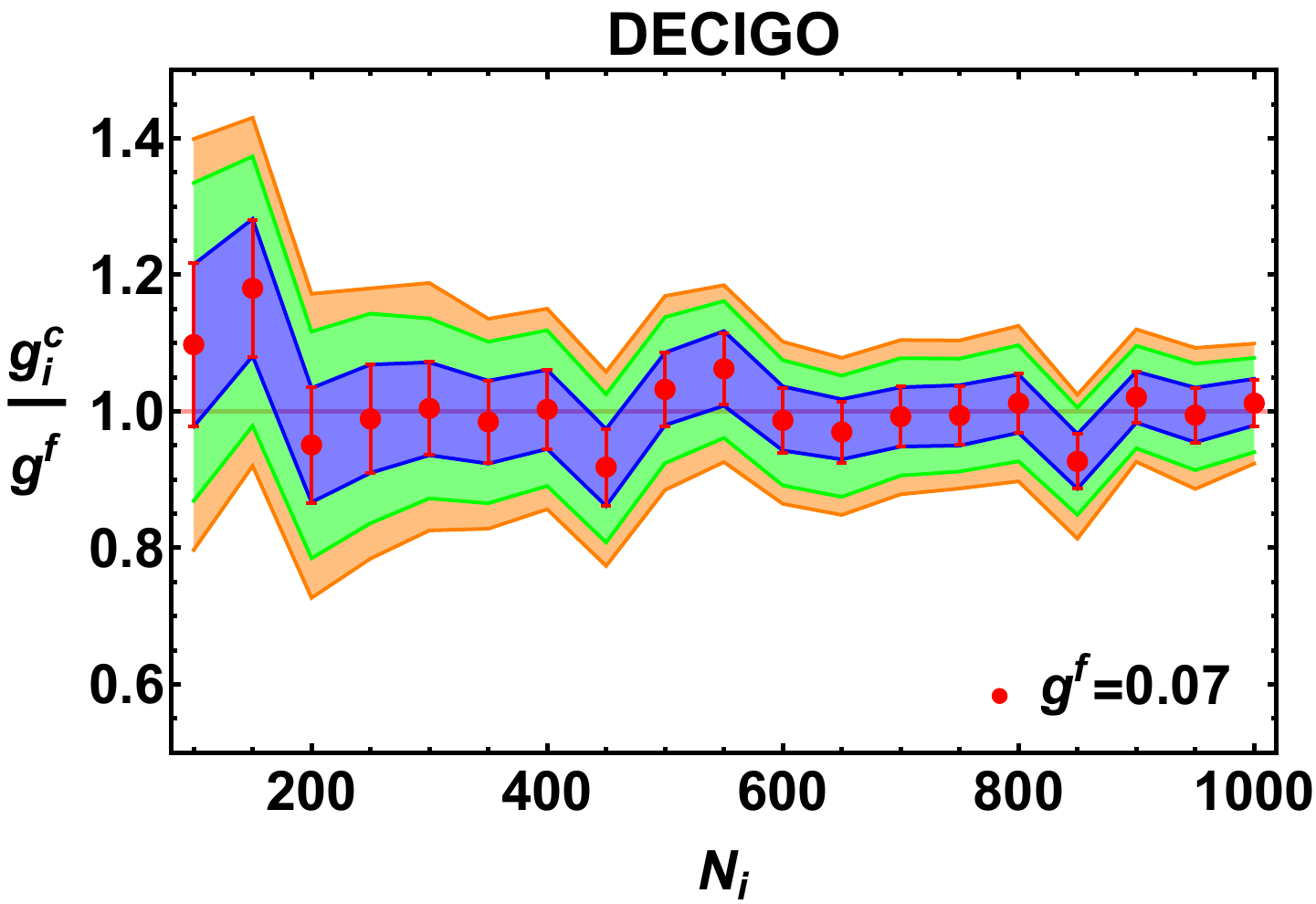}
  \includegraphics[width=0.32\textwidth]{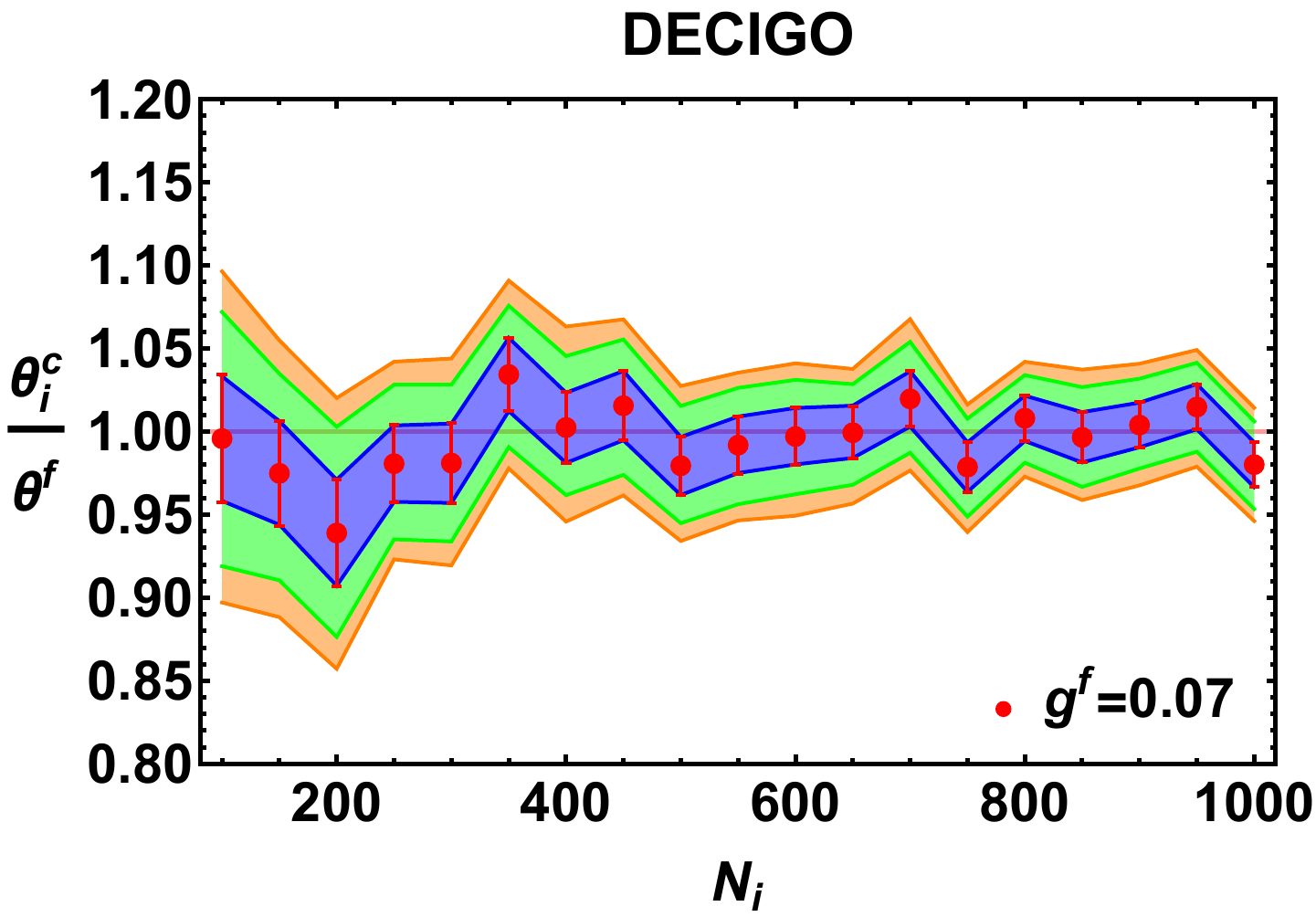}
  \includegraphics[width=0.32\textwidth]{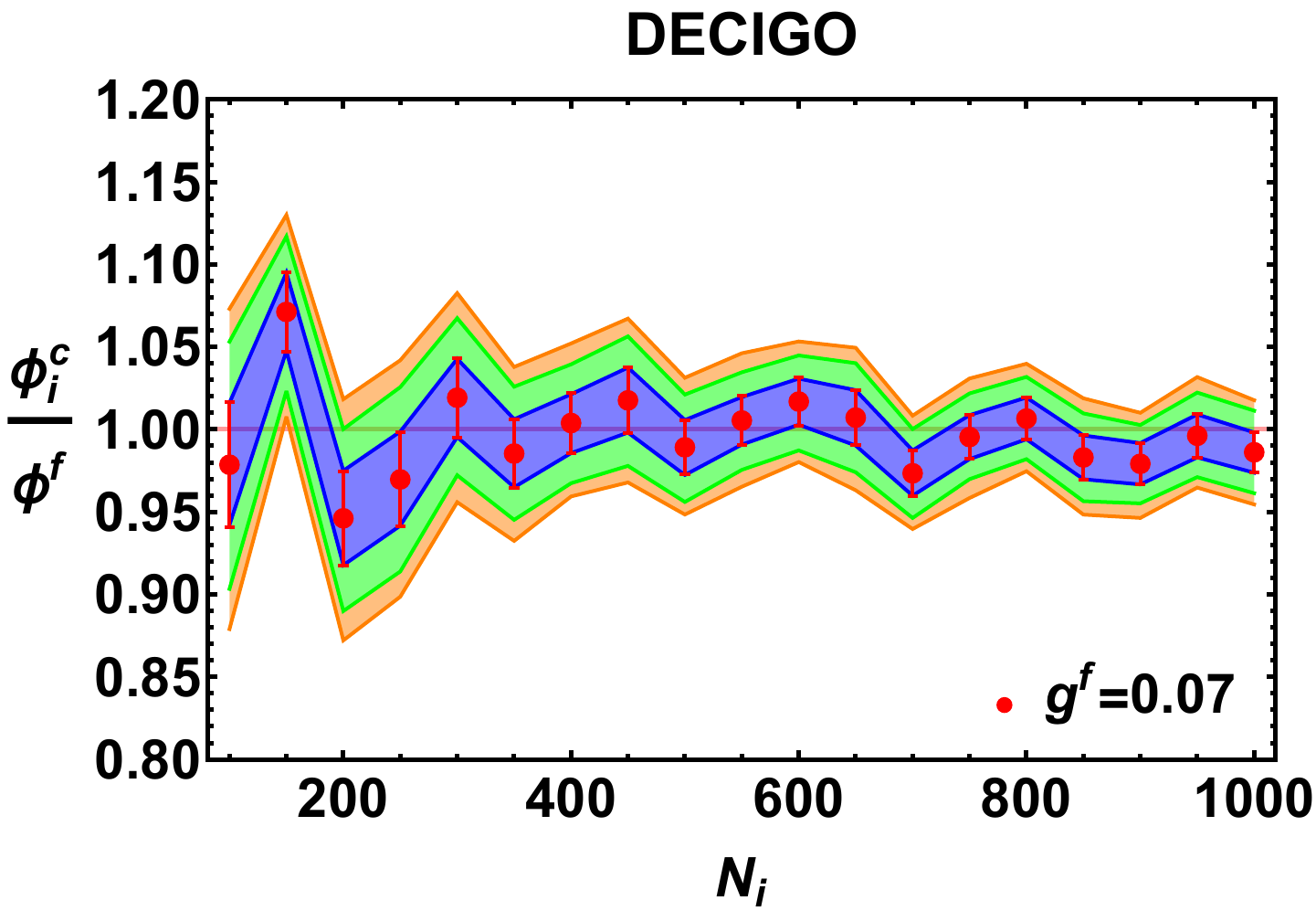}\\
  \caption{The constraint ability of DECIGO with respect to the varying number of standard siren events for given fiducial values of $g^f=0.03$ (top panels) and $g^f=0.07$ (bottom panels). The best constrained values divided by the corresponding fiducial values for $g$ (left panels), $\theta$ (medium panels) and $\phi$ (right panels) are labeled by the red dots with standard deviation error bars. The blue/green/orange shaded regions are of 1$\sigma$/2$\sigma$/3$\sigma$ C.L., respectively.}\label{fig:DECIGOSigma}
\end{figure*}

\section{Summary of results}\label{sec:results}

Using the approach introduced in \ref{sec:simulations}, we first obtain the result for LISA, shown in Fig. \ref{fig:LISA}. The cosmic isotropy can be ruled out at $3\sigma$ C.L. and the dipole direction can be constrained roughly around $20\%$ at $2\sigma$ C.L., as long as the dipole amplitude is larger than $0.03$, $0.06$ and $0.025$ for MBHB models Q3d, pop III and Q3nod with increasing constraining ability, respectively, which is consistent with the results from \cite{Cai:2017yww}.

Next, we plot the standard deviations $(\sigma_{g^c}, \sigma_{\theta^c}, \sigma_{\phi^c})$ normalized by corresponding fiducial values $(g^f, \theta^f, \phi^f)$ in logarithmic units with respect to the fiducial dipole amplitude $g^f$ and GW event number $N$, which are shown in Fig. \ref{fig:ETandDECIGO}. With the increase of $g^f$ and $N$, the three standard deviations all get smaller and smaller, indicating the improvement of constraint ability. However, this improvement for $g$ is not as significant as that for $\theta\ \rm and\ \phi$. While for DECIGO, they share almost the same tendency. It can be obviously seen that DECIGO has better performance than ET, comparing the top panels with the bottom panels.

In Fig. \ref{fig:ETSigma}, we choose some representative panels by fixing the fiducial value of amplitude $g^f$ (shown in the bottom of each panel), to explicitly illustrate how well ET can put constraint on anisotropy with respect to the given number $N$ of GW events. We find that once the dipole amplitude $g^f$ is increased to $0.06$, the cosmic isotropy can be ruled out at $3\sigma$ C.L. with no less than $200$ GW events, and the dipole direction can be constrained within $20\%$ at $3\sigma$ C.L. if $g^f$ is near $0.1$.

The similar figure for DECIGO is presented in Fig. \ref{fig:DECIGOSigma}, but with different values of amplitude $g^f$. From the top panels, we can see that fewer GW events is needed than that of ET case, namely $N\gtrsim100$, in order to rule out the cosmic isotropy at $3\sigma$ C.L. as long as $g\gtrsim0.03$. Speaking of the constraint on the dipole direction, with $g^f\gtrsim0.07$ and $N\gtrsim100$, we can constrain it within $10\%$ at $3\sigma$ C.L., much better than ET does. However, the constraint ability from LISA with 30 standard siren events is roughly comparable with those from ET and DECIGO with few hundreds of standard siren events.

\section{Conclusions}\label{sec:conclusion}

In this paper, we use GW as the standard siren to investigate the constraint ability on the anisotropy in the Universe expansion. Comparing with the approach using SNe Ia data sets, GW enjoys the advantage of high accuracy with less sources of systematic errors, since the luminosity distance can be inferred directly and precisely from the gravitational waveform of coalescing binaries. Besides, the redshift can also be determined by the accompanying EM counterparts. With presumed dipole anisotropy, we construct the simulated data of GW events from BNS and BHNS for both ET and DECIGO and GW events from MBHB for LISA as well.

For LISA, we find that the cosmic isotropy can be ruled out at $3\sigma$ C.L. so long as the dipole amplitude is larger than $0.03$, $0.06$ and $0.025$ for MBHB models Q3d, pop III and Q3nod, respectively. At the same time, the dipole direction can be constrained roughly around $20\%$ at $2\sigma$ C.L.. For ET with no less than 200 GW events, we can rule out the cosmic isotropy at $3\sigma$ C.L. if the dipole amplitude is larger than $0.06$, and the dipole direction can be constrained within $20\%$ at $3\sigma$ C.L. if the dipole amplitude is close to $0.1$. For DECIGO with no-less than 100 GW events, the cosmic isotropy can be ruled out at $3\sigma$ C.L. for dipole amplitude larger than 0.03, and the dipole direction can even be constrained within $10\%$ at $3\sigma$ C.L. if dipole amplitude is larger than 0.07. Our work manifests the promising perspective of the constraint ability on the cosmic anisotropy from the standard siren approach.

\begin{acknowledgments}
We want to thank Zong-Kuan Guo, Bin Hu, Li-Wei Ji and Wu-Tao Xu for helpful discussions. S. J. W. would like to thank David Weir and Kari Rummukainen for the warm hospitality during the visit at Helsinki Institute of Physics at the final stage of this paper. S. J. W. also wants to thank Nicola Tamanini for stimulating discussions and useful correspondence. This work is supported by the National Natural Science Foundation of China Grants No. 11690022, No. 11435006 and No. 11647601, and by the Strategic Priority Research Program of CAS Grant No. XDB23030100 and by the Key Research Program of Frontier Sciences of CAS. Y. T. is supported by the National Natural Science Foundation of China Grant No. 210100088.
\end{acknowledgments}

\bibliographystyle{utphys}
\bibliography{ref}

\end{document}